\documentclass[preprint,authoryear,11pt]{elsarticle}  
\usepackage{fullpage}
\usepackage{graphicx}
\usepackage{amssymb}
\usepackage{amsmath}
\usepackage{multirow}
\usepackage{lineno}
\usepackage{subfig}

\usepackage{color}
\usepackage{soul}
\usepackage{array}
\usepackage{hyperref}
\usepackage[colorinlistoftodos]{todonotes}
\usepackage[]{algorithm2e}
\usepackage{varwidth}
\graphicspath{{./figs-rui/}}

\newcommand{\specialcell}[2][c]{%
  \begin{tabular}[#1]{@{}c@{}}#2\end{tabular}}

\newcommand{\xmb}[1]{\ensuremath{\mathbf{#1}}}
\newcommand{\xmbs}[1]{\ensuremath{\boldsymbol{#1}}}

\journal{}

\begin{document}

\begin{frontmatter}

  \title{Realistic Representation of Grain Shapes in CFD--DEM Simulations of Sediment Transport: A
    Bonded-Sphere Approach}

 \author{Rui Sun} \ead{sunrui@vt.edu}
 \author{Heng Xiao\corref{corxh}} \ead{hengxiao@vt.edu}
 \address{Department of Aerospace and Ocean Engineering, Virginia Tech, Blacksburg, VA 24060, United
 States}

 \cortext[corxh]{Corresponding author. Tel: +1 540 231 0926}

\begin{abstract}
  Development of algorithms and growth of computational resources in the past decades have enabled
  simulations of sediment transport processes with unprecedented fidelities. The Computational Fluid
  Dynamics--Discrete Element Method (CFD--DEM) is one of the high-fidelity approaches, where the
  motions of and collisions among the sediment grains as well as their interactions with surrounding
  fluids are resolved.  In most DEM solvers the particles are modeled as soft spheres due to
  computational efficiency and implementation complexity considerations, although natural sediments
  are usually mixture of non-spherical (e.g., disk-, blade-, and rod-shaped) particles. Previous
  attempts to extend sphere-based DEM to treat irregular particles neglected fluid-induced torques
  on particles, and the method lacked flexibility to handle sediments with an arbitrary mixture of
  particle shapes. In this contribution we proposed a simple, efficient approach to represent common
  sediment grain shapes with bonded spheres, where the fluid forces are computed and applied on each
  sphere. The proposed approach overcomes the aforementioned limitations of existing methods and has
  improved efficiency and flexibility over existing approaches.  We use numerical simulations to
  demonstrate the merits and capability of the proposed method in predicting the falling
  characteristics, terminal velocity, threshold of incipient motion, and transport rate of natural
  sediments. The simulations show that the proposed method is a promising approach for faithful
  representation of natural sediment, which leads to accurate simulations of their transport
  dynamics.  While this work focuses on non-cohesive sediments, the proposed method also opens the
  possibility for first-principle-based simulations of the flocculation and sedimentation dynamics
  of cohesive sediments. Elucidation of these physical mechanisms can provide much needed
  improvement on the prediction capability and physical understanding of muddy coast dynamics.
\end{abstract}


\end{frontmatter}


\section{Introduction}
\label{sec:intro}

Computational Fluid Dynamics--Discrete Element Method (CFD--DEM) has been increasingly used in the
study of sediment transport~\citep{schmeeckle14ns,sun16awr}. In this method, individual particles
are tracked in a Lagrangian framework, which offers more insight compared to continuum-based
descriptions of the particle phase.  In most standard DEM solvers the particles are represented as
soft spheres, i.e., a particle is characterized by a few geometric and mechanical constants (e.g.,
diameter, Young's modulus, and restitution coefficient, and friction coefficient)~\citep{lammps}.
Representing particles as spheres immensely simplifies the kinematic description and the contact
detection, which are needed for the calculation of particle--particle interaction forces.
Specifically, the kinematic state of a spherical particle can be uniquely determined by its center
location as well as its translational and angular velocities, while its orientation does not need to
be described. The contact and overlap between two particles can be straightforwardly computed from
the difference between their center separation distance and the sum of their radii. The simple
kinematic description and contact detection offered by the spherical particle models enable modern
DEM solvers to simulate systems up to $10^7$ particles on computers with a few hundred
cores~\citep{guo2012numerical,sun2016sedi}.  Despite the success of the sphere-based representations
in particle-laden flows in a wide range of applications, researchers have recognized its limitations
in scenarios where irregular particle shapes do play a role. In the past decade, researchers have
pursued realistic representation of particle geometry and contact force modeling.  However, most of
the efforts so far have focused on DEM of dry granular flows, and the literature on using irregular
particles in CFD--DEM simulations is relatively sparse.  In this section we review the literature in
both fields and propose a bonded-sphere representation for representing irregular particles for
CFD--DEM simulations with particular emphasis on sediment transport applications.

\subsection{Irregular Particles in DEM Simulations of Dry Granular Flows}

A number of studies have used DEM with non-spherical particles to study dry granular flows.  For
example, DEM based on general polygons was used in geotechnical engineering to study rock fraction
and sand production~\citep{connor97}.  It can be expected that such a generic description of
particle shape is challenging to implement, particularly for three-dimensional simulations.
Moreover, it is computationally expensive for contact detection, particularly for systems with a
large number of particles.  The computational cost was not likely a major concern for rock mechanics
applications, since the system dynamics is often dominated by a small number of large particles.
However, this is not the case in sediment transport applications, where the number of particles is
much larger, and polygon-based representation of particles is likely to be infeasible in DEM
simulations of sediment transport.  Recently, DEM solvers based on non-spherical particle models
such as disks, cylinders (rods), and ellipsoids, or superellipsoids have been developed and used to
study the macro- and micro-mechanical behaviors of the dry assemblies of these
particles~\citep{kodam2010cylindrical-a,kodam2010cylindrical-b,guo2012numerical,guo2013granular}.
For example, \citet{guo2012numerical} studied the dry assembly of rod-like particles in simple shear
flows.  Their simulation results suggest that the particle shape has dramatic influences on the
normalized shear stress, collision rate, and particle orientations~\citep{guo2012numerical}. Since
natural sediment particles are rarely spherical, we need to reexamine the soft sphere particle
models that are commonly used in current CFD--DEM solvers for sediment transport simulations.
However, using a specific non-spherical particle shape (disk or rod) in a DEM simulation may not be
sufficient for a realistic presentation of natural sediments, which are rarely monodispersed but are
often a mixture of a wide range of shapes. For example, the calcareous sediment sample collected on
the beach of Oahu, Hawaii, and used in the study of \citet{smith03sc,smith04im} consists of 43\%
disk-shaped, 9\% blade-shaped, 14\% rod-like, and 34\% equant (equidimensional) particles based on
the Zingg classification~\citep{zingg1935beitrag,krumbein1941measurement}. Mixing the elementary
shapes (disks, rods, and spheres) would make the contact detection even more difficult than a
homogeneous system (e.g., with rods only or disks only).

One alternative to represent irregular, non-spherical particles is by using several bonded spheres,
possibly with overlapping.  The spheres that are bonded to form the irregular particles are referred
to as \emph{component spheres}~\citep{favier1999shape,price2007sphere}.  In contrast, hereafter we
will use \emph{particles} to refer to the composite particle consisting of one or more component
spheres.  Previous studies suggest that increasing the number of component spheres used to represent
each irregular particle improves geometric accuracy of the representation but can cause the force
modeling accuracy to deteriorate.  \citet{kruggel2008study} examined the validity of the
bonded-spheres approach in DEM simulations of dry granular flows by studying the trajectories of a
spherical particle during collisions with wall and with other particles. The results obtained with a
single sphere representation and those with many bonded spheres were compared. They concluded that
the bonded-sphere representation is an advancement compared to the spherical bodies but also pointed
out the approximation nature of the representation. However, further studies are needed to make more
general conclusions on the validity of the bonded-sphere representation in the context of large
particle assembly, which is likely to be challenging.  \citet{guo2012numerical} used bonded-spheres
to represent cylindrical particles and examined the shear stresses and collision rate in a dry
granular flow subject to simple shearing flow (Couette type flow).  The results from the
bonded-sphere representation with those from actual cylinders were compared. They found that the two
representations are almost identical in dilute flows with volume fractions $\varepsilon_s < 0.1$,
which are dominated by particle movement. In dense flows with $\varepsilon_s > 0.2$, the two
representations lead to different results, likely caused by the bumpy surface bonded-spheres,
particularly when the friction is not considered.

The studies reviewed above focused on the geometric accuracy of the bonded-sphere representation,
which improves as more component spheres are used to represent a particle, albeit at increased
computational costs.  Overlapping several particle can improve geometric representation of irregular
particles~\citep[e.g.,][]{guo2012numerical} by reducing the surface bumpiness.  On the other hand,
\citet{kodam2009force} demonstrated that the accuracy of force modeling also needs to be accounted
for when using the sphere derived forces in describing the collisions among irregular
particles. They concluded that the inaccuracy in force modeling can be reduced by using fewer
component spheres in representing a composite particle. This requirement, however, conflicts with
geometrical accuracy of the bonded-sphere representation, which favors using more component spheres.

\subsection{Irregular Particles in CFD--DEM Simulations of Particle-Laden Flows}

Despite the large volume of literature of representing irregular particles with bonded spheres in
DEM simulations of dry assembly, using irregular particles in CFD--DEM simulations of particle-laden
flows is relatively rare. An notable example is the work of \citet{calantoni2004modelling}, who used
such a representation in sediment transport simulations. One of challenges for using irregular
particles in CFD--DEM simulations is that one needs to consider not only the accuracy of geometry
and force modeling, but also the accuracy in fluid--particle interactions forces. It has been
mentioned above that in dry DEM simulations the geometry and force modeling accuracy already pose
conflicting requirements by preferring larger and smaller number of component spheres,
respectively. The consideration of fluid--particle interaction further complicates the picture.

\begin{figure}[htbp]
  \centering
  \includegraphics[width=0.55\textwidth]{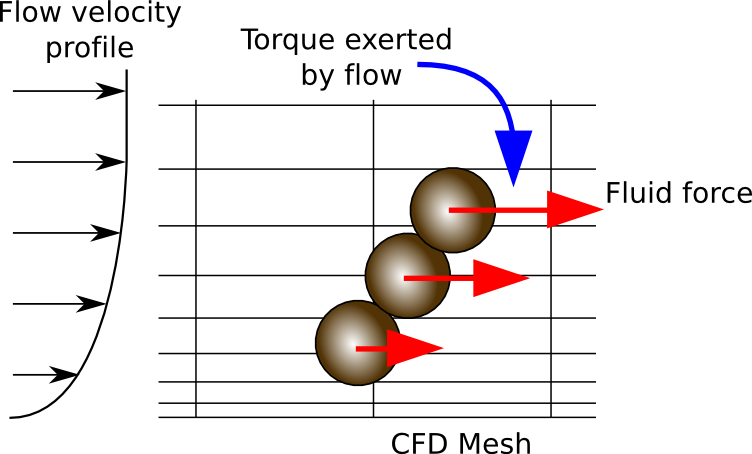}
  \caption{Illustration of a rod-shaped particle consisting of three bonded component particles
    experiencing shear flow at the boundary layer. The red arrows indicate the magnitude and
    direction of the fluid force exerted on the each sphere particle; the blue arrow indicates the
    direction of the torque exerted on the bonded particle by the fluid force.  A typical mesh used
    in CFD--DEM simulations is plotted using solid lines.}
  \label{fig:mean-shear}
\end{figure}

In the work of \citet{calantoni2004modelling}, they bonded two spheres of different yet specified
diameters to represent a dumb-bell shaped irregular particle that is representative of natural
sediments. Hence, the particles in their simulations are irregular but monodispersed, consisting of
particle with the same dumb-bell shape. The center of mass (CM) and the momentum of inertia are
calculated analytically in advance and are subsequently used in the DEM simulations. The fluid
forces on the irregular particles are calculated based on a spherical particle of the same mass, and
the fluid forces are applied at the center of mass of the irregular particle. Therefore, the torques
exerted by the fluid on the irregular particles are not considered. In flows with strong velocity
gradients (e.g., in boundary layers), which are typical in sediment transport, different parts
(i.e., component spheres) of the particle may be surrounded by fluids of different velocities, and
thus the torque exerted by the shearing of the flow can play a significant role. This effect is
particularly pronounced for particles of large aspect ratios, e.g., rods and blades.  Noted that
neglecting the fluid torque on the particle in \citet{calantoni2004modelling} is consistent with the
overall resolution of the CFD--DEM coupling. Specifically, the spatially averaged Navier--Stokes
equations as a basis of CFD--DEM rely on the assumption that the representative volume and the CFD
mesh cells are much larger than the particle diameters. Consequently, it is unlikely that different
parts of a particle would experience different fluid velocities. In other words, the spatial
variation $\Delta u = d_p \frac{\partial U_f}{\partial x}$ of fluid velocity $U_f$, or equivalently
the mean shear rate experienced by the particle over the length scale of a particle diameter $d_p$,
is negligible.  However, recent development in CFD--DEM has partly alleviated the CFD cell size
constraints. By applying a diffusion kernel in the averaging of particle data to Eulerian fields
\citep{Capecelatro_13_AE,sun14db1}, it is now possible to use CFD mesh with cells sizes much smaller
than particle diameter while still yielding mesh-independent results.  Consequently, the
diffusion-based coarse-graining procedure makes it possible to have particles span over several CFD
cells and to experience significant mean flow shearing, which is depicted in
Fig.~\ref{fig:mean-shear}.  Consistent with the development of this new flexible averaging
procedure, it is preferred to develop a bonded-sphere approach where both particle contact forces
and fluid--particle interactions forces are computed based on individual component spheres.

\subsection{Overview and Merits of the Proposed Method}


In this work we propose a method to represent sediment particles of arbitrary shapes in CFD--DEM
simulations by using bonded spheres. The representation attempts to strike a balance among
conflicting requirements of accurate representing particle geometry (i.e., aspect ratio, particle
mass, volume, and momentum of inertia), collision forces, and fluid--particle interactions.  To this
end, the fluid--particle interaction forces are computed and applied to individual component spheres
and not on the center of mass of the composite particle as in~\citet{calantoni2004modelling}.  The
sphere-based computation of fluid--particle interaction forces offers better flexibility, the
ability to compute particle-level torque, and potentially better accuracy. In particular, it avoids
using ad hoc correction methods to estimate the forces on the entire particle.  Moreover, we require
all component spheres have a wetted surface, since it is difficult to justify the use of
sphere-based fluid--particle interaction forces. Finally, particle overlapping is avoided as it
would cause difficulties in computing fluid--particle interactions forces.

The objective of this work is to demonstrate that the proposed method is capable of simulating the
most important qualitative features and quantitative integral quantities that are critical for
sediment transport, including falling trajectory characteristics, terminal velocity, incipient
motion, and transport rate. The work of \citet{calantoni2004modelling} has shown that, by
representing an irregular shaped natural sediment particle with bonded-spheres, DEM simulations are
able to reproduce the experimentally observed repose angle of dry sand.  Compared to the approach of
\citet{calantoni2004modelling} and other existing approaches for representing irregular
particles~\citep{kodam2010cylindrical-a,kodam2010cylindrical-b,guo2012numerical,guo2013granular}, the
advantages of the proposed method are detailed below.
\begin{enumerate}
\item By applying fluid particle forces to individual particles, the current method enables a better
  representation of fluid exerted torque on the composite particles. 
\item The proposed method can utilize the widely validated empirical formulas for fluid forces
  (e.g., drag laws~\citep{wen1966mechanics,mfix93,di1994voidage}) on spherical
  particles, since only the fluid forces on spherical particles are need. In contrast, when
  representing particles with cylinder or ellipsoids, one must rely on empirical corrections to
  account for the shapes, which are often based on shape factors. Studies of drag laws for
  non-spherical particles are relatively sparse in the literature, and no consensus exists so far.
  Using these empirical corrections can lead to large uncertainties. The lack of experience for
  empirical correlations is even more acute for other forces (e.g., lift).
\item The generic, simplified representation of particles of arbitrary shapes allows the flexibility
  of treating sediments consisting of particles with different shapes and sizes, which is typical of
  natural sediments. Specifically, analytical calculations of center of mass and moment of inertia
  as performed by \citet{calantoni2004modelling} are not needed. Furthermore, detection of particle
  collisions is simplified with the bonded-sphere, since we only need to detect collisions and
  computing forces among spheres, which is much simpler than those for other irregular shapes
  \citep[e.g.,][]{kodam2010cylindrical-a,guo2012numerical}.  Consequently, even though the bonded-sphere
  representation may lead to larger number of particles, the overall efficiently is like to be
  higher than the representation with non-spherical shapes.
\item The proposed method enables efficient treatment of breakup and agglomeration of particles
  assembly.  This capability paves way for the CFD--DEM simulation of cohesive sediments and would
  enable the CFD--DEM as a powerful tool for the study of the flocculation dynamics in
  sedimentation and transport of cohesive sediments.
\end{enumerate}

The rest of the paper is organized as follows. The methodology of the present model is introduced in
Section~\ref{sec:method}, including the mathematical formulation of fluid equations, the particle
motion equations, the fluid--particle interactions, and the geometry and dynamics of irregular
particles. In Section~\ref{sec:veri}, the a priori test of the properties of irregular particle is
detailed. In Section~\ref{sec:vali}, the results obtained in the validation tests are discussed to
demonstrate the present approach is capable of modeling the rolling, the incipient motion, and
sediment transport of irregular particles. Finally, Section~\ref{sec:conclusion} concludes the
paper.

\section{CFD--DEM of Flows Laden with Irregular Particles}
\label{sec:method}

In traditional CFD--DEM the particles are modeled as soft spheres. The velocity and location of each
particle are resolved, from which the interparticle contact and the deformation of each particle are
derived~\citep{cundall79}. The interaction forces between a particle and the fluid are modeled based
on empirical correlations~\citep{tsuji93}.  The fluid flows are described by using a locally averaged
Navier--Stokes equation with the spatially averaged effects of the particles accounted for.  In this
work we extend the traditional CFD--DEM to simulate fluid flows laden with irregular particles. Each
particle consists of one or more component spheres bonded together.  The contact forces between
component spheres and the interactions between the spheres and the fluid are computed in the same
way as in traditional CFD--DEM, but the component spheres forming a particle are moved in such a way
that the particle move in a rigid body motion. That is, their relative positions do not change. In
this section, equations describing the fluid flows and the particle motions are first presented and the
models for the fluid--particle interaction forces are then introduced. Finally, the algorithm to
represent sediment particles of arbitrary shape is presented.

\subsection{Locally-Averaged Navier--Stokes Equations for Fluid Flows}

The fluid phase is described by the locally-averaged incompressible Navier--Stokes equations.
Assuming constant fluid density \(\rho_f\), the governing equations for the fluid
are~\citep{anderson67,kafui02}:
\begin{subequations}
 \label{eq:NS}
 \begin{align}
  \nabla \cdot \left(\varepsilon_s \xmb{U}_s + {\varepsilon_f \xmb{U}_f}\right) &
  = 0 , \label{eq:NS-cont} \\
  \frac{\partial \left(\varepsilon_f \xmb{U}_f \right)}{\partial t} + \nabla \cdot
  \left(\varepsilon_f \xmb{U}_f \xmb{U}_f\right) &
  = \frac{1}{\rho_f} \left( - \nabla p + \varepsilon_f \nabla \cdot \xmbs{\mathcal{R}} +
  \varepsilon_f \rho_f \xmb{g} + \xmb{F}^{fp}\right), \label{eq:NS-mom} 
 \end{align}
\end{subequations}
where \(\varepsilon_s\) is the solid volume fraction; \( \varepsilon_f = 1 - \varepsilon_s \) is the
fluid volume fraction; \( \xmb{U}_f \) is the fluid velocity. The terms on the right hand side of
the momentum equation are: pressure (\(p\)) gradient, divergence of the stress tensor \(
\xmbs{\mathcal{R}} \) (including viscous and Reynolds stresses), gravity, and fluid--particle
interactions forces, respectively. In the present study, we used large-eddy simulation to resolve
the flow turbulence in the computational domain. We applied the one-equation eddy viscosity model
proposed by~\cite{yoshizawa85sd} as the sub-grid scale (SGS) model. The Eulerian fields
$\varepsilon_s$, $\xmb{U}_s$, and $\xmb{F}^{fp}$ in Eq.~(\ref{eq:NS}) are obtained by averaging the
information of Lagrangian particles.

\subsection{Discrete Element Method for Particles}

As in the CFD--DEM approach, the translational and rotational motion of each component sphere is
calculated based on Newton's second law as the following equations~\citep{cundall79,ball97si}:
\begin{subequations}
 \label{eq:newton}
 \begin{align}
  m \frac{d\xmb{u}}{dt} &
  = \xmb{f}^{pp}  + \xmb{f}^{fp} + m \xmb{g} \label{eq:newton-v}, \\
  I \frac{d\xmbs{\Psi}}{dt} &
  = \xmb{T}^{pp} + \xmb{T}^{fp} \label{eq:newton-w},
 \end{align}
\end{subequations}
where the superscript $f$ indicates fluid phase and $p$ indicates the component spheres within the
composite particles; \( \xmb{u} \) is the velocity of the particle; $t$ is time; $m$ is particle
mass; \(\xmb{f}^{pp} \) represents the contact forces due to interparticle collisions and the
bonding between component spheres; \(\xmb{f}^{fp}\) denotes fluid--particle interaction forces
(e.g., drag, lift force and buoyancy); \(\xmb{g}\) denotes body force. Similarly, \(I\) and
\(\xmbs{\Psi}\) are angular moment of inertia and angular velocity, respectively, of the particle;
\(\xmb{T}^{pp}\) and \(\xmb{T}^{fp}\) are the torques due to particle--particle collisions or
bonding and fluid--particle interactions, respectively.  To compute the collision forces and
torques, the component particles are modeled as soft spheres with inter-particle contact represented
by an elastic spring and a viscous dashpot.  Further details can be found in~\cite{tsuji93}.

\subsection{Fluid--Particle Coupling}

The fluid--particle interaction force \(\xmb{F}^{fp}\) consists of buoyancy \( \xmb{F}^{buoy} \),
drag \( \xmb{F}^{drag} \), and lift force \(\xmb{F}^{lift}\). The drag on an individual component
sphere $i$ is formulated as:
\begin{equation}
  \mathbf{f}^{drag}_i = \frac{V_{p,i}}{\varepsilon_{f, i} \varepsilon_{s, i}} \beta_i \left( \mathbf{u}_{p,i} -
  \mathbf{U}_{f, i} \right),
  \label{eqn:particleDrag}
\end{equation}
where \( V_{p, i} \) and \( \mathbf{u}_{p, i} \) are the volume and the velocity of particle $i$,
respectively; \( \mathbf{U}_{f, i} \) is the fluid velocity interpolated to the center of particle
$i$; \( \beta_{i} \) is the drag correlation coefficient which accounts for the presence of other
particles. The $\beta_i$ value in the present study is based on~\cite{mfix93}:
\begin{equation}
\beta_i = \frac{3}{4}\frac{C_d}{V^2_r}\frac{\rho_f |\xmbs{u_{p,i}} - \xmbs{U_{f,i}}|}{d_{p,i}}
\mathrm{, \quad with \quad} C_d = \left( 0.63+0.48\sqrt{V_r/\mathrm{Re_p}} \right),
  \label{eqn:beta-i}
\end{equation}
where the particle Reynolds number $Re_p$ is defined as:
\begin{equation}
  \mathrm{Re_p} = \rho_i d_{p,i} |\xmbs{u_{p,i}} - \xmbs{U_{f,i}}|;
  \label{eqn:p-re}
\end{equation}
the $V_r$ is the correlation term:
\begin{equation}
  V_r = 0.5\left( A_1 - 0.06\mathrm{Re_p}+\sqrt{(0.06\mathrm{Re_p})^2 +
  0.12\mathrm{Re_p}(2A_2 - A_1)+A_1^2} \right),
  \label{eqn:drag-vr}
\end{equation}
with
\begin{equation}
  A_1 = \varepsilon_f^{4.14}, \quad
  A_2 =
  \begin{cases}
    0.8\varepsilon_f^{1.28} & \quad \text{if } \varepsilon_f \le 0.85, \\
    \varepsilon_f^{2.65}    & \quad \text{if } \varepsilon_f > 0.85.\\
  \end{cases}
  \label{eqn:drag-A}
\end{equation}
In addition to drag, the lift force on a spherical component particle is modeled
as~\citep{saffman65th,rijn84se1}:
\begin{equation}
  \mathbf{f}_{i}^{lift} = C_{l} \rho_f \nu^{0.5} d_{p,i}^{2} \left( \mathbf{u}_{p,i} - \mathbf{U}_{f,i}
  \right) \boldsymbol{\times} \nabla \mathbf{U}_{f,i},
  \label{eqn:particleLift}
  \end{equation}
where $\boldsymbol{\times}$ indicates the cross product between vectors and tensors, and $C_{l} =
1.6$ is the lift coefficient.

\subsection{Geometry and Dynamics of Irregular Particles}
\label{sec:method-irr}

As described above, each particle is represented with a number of component spheres, and
the motion of each component sphere and the fluid forces exerted thereon are computed individually.
Each component sphere may experience deformation through contacts with component spheres
representing other particles. However, the component spheres forming the same particle preserve
their relative positions in the entire simulation.  This is ensured by the following procedure for each particle at
every integration time step:
\begin{enumerate}
\item Compute the total forces and torques on the irregular particle by summing up the forces and
  torques exerted by other particles on each component spheres forming the
  particle,
\item Compute the position and orientation of the irregular particle, whose center of mass and momentum
  of inertia are computed beforehand (straightforward for non-overlapping spheres), and
\item Update the position and orientation of all component spheres in the particle.
\end{enumerate}
With this procedure, the particles experience only rigid-body translations and rotations.  For
accuracy considerations in  computing the fluid--particle interaction forces, the component spheres
in the same particle do not overlap initially or throughout the simulation, since their relative
positions do not change. Consequently, there are no contact forces among these spheres. However,
even if they are allowed to overlap, it would not affect the computation of particle motions
described above.

We chose to use as few component spheres as possible to represent the irregular particles. A notable
difference in CFD--DEM simulations of sediment transport compared to the dry granular flow and
particle-laden flows in industrial processes is that the particle shapes are not well characterized
and highly heterogeneous. Often, only a few parameters such as the sieving diameter and nonuniform
coefficient are available. This is different from the tablet modeling in pharmaceutical industry,
where the shape of the tablet is known precisely~\citep{song2006contact,kodam2012discrete}.
Therefore, it is not essential to represent the detailed particle geometry, since there exists large
natural variations and the uncertainties in the characterization. 

\begin{figure}[htbp]
  \centering
  \includegraphics[width=0.55\textwidth]{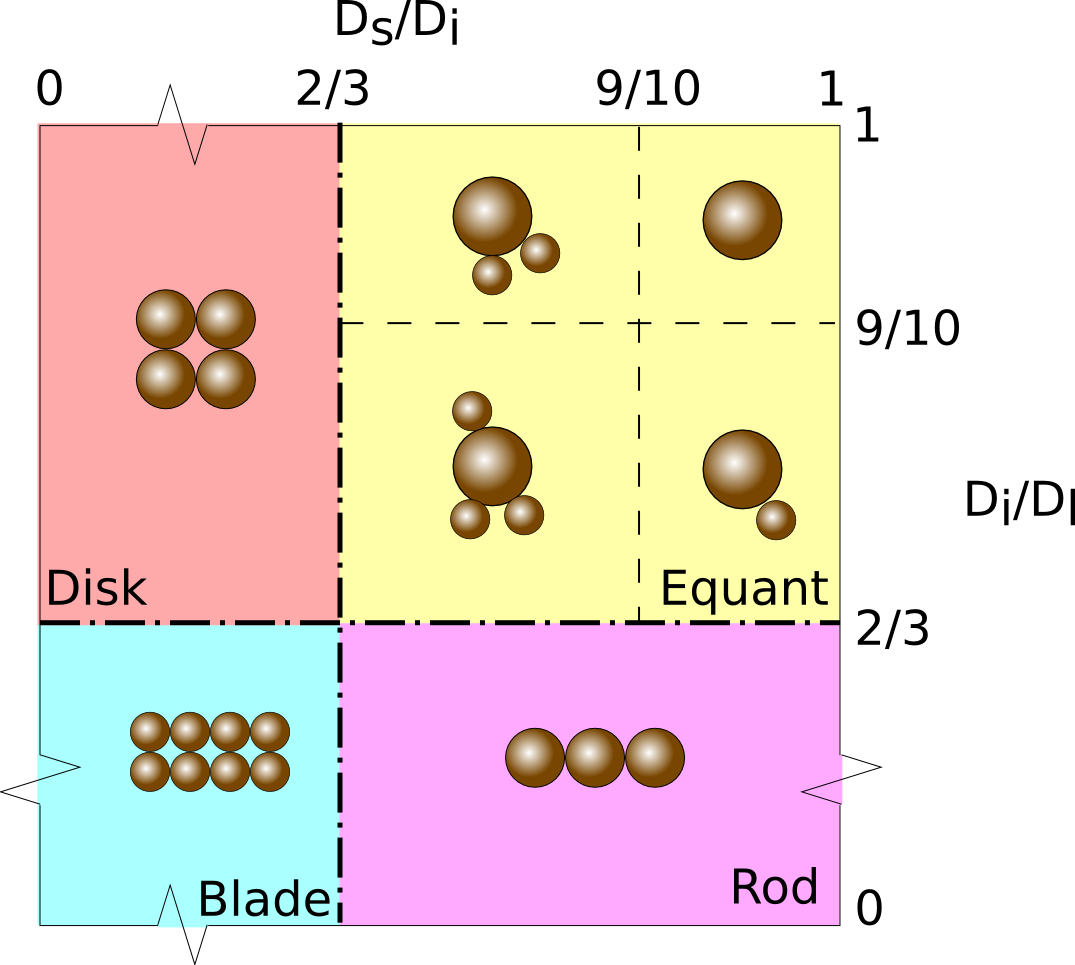}
  \caption{Particle shape classification by $D_i/D_l$ and $D_s/D_i$ based on the Zingg’s diagram}
  \label{fig:layout}
\end{figure}

It is a daunting task even to characterize a generic irregular particle with a few parameters. The
following parameters are frequently included: sphericity, Corey shape factor, and Zingg's
classification. The sphericity is defined as the ratio between the sphere having the same volume as
the particle and the actual surface area of the particle. The Corey shape factor (and many other
shape factors) and the Zingg's classification are based on the ratios among the long, intermediate,
and short axes ($D_l$, $D_i$, and $D_s$, respectively) in~\cite{smith03ef}.  However, any finite set
of these parameters is not able to provide a complete description of particle shape. Fortunately,
most natural sediment grains do exhibit relatively simple shapes as they have been worked for many
years. As a result, the ratios $D_i/D_l$ and $D_s/D_i$ provide a rather good description of most
sediment particles. Based on the Zingg's diagram in Fig.~\ref{fig:layout}, they can be classified to
disk-, blade-, rod-, and equant shaped particles. In the proposed work we take a simplistic approach
to use spheres to represent each of the four classes of particles as described below.

\begin{figure}[htbp]
  \centering
  \includegraphics[width=0.95\textwidth]{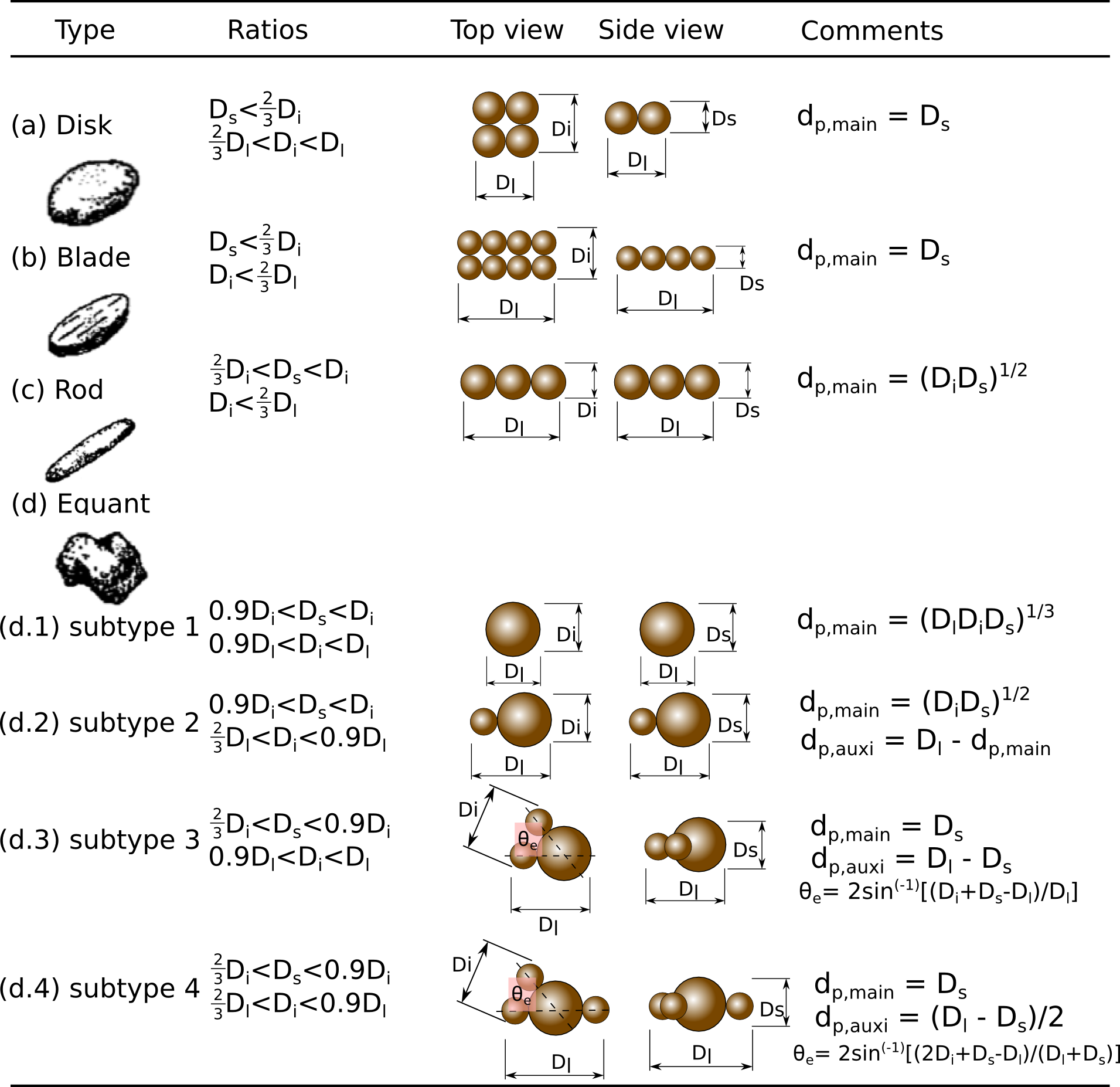}
  \caption{Configuration of the ``discretization'' of irregular particles. The pictures of the
    irregular sediment particles are from the Ph.D. thesis of~\cite{demir00if} downloaded from the
    Internet resource of the Department of Geological Sciences of University of Saskatchewan.}
  \label{fig:layout-table}
\end{figure}

\begin{itemize}
\item Disk-shaped particles are those whose long and intermediate axes are of similar dimension
  ($\frac{2}{3}D_l < D_i < D_l$), and the short axis is much shorter, i.e., $D_s < \frac{2}{3} D_i$.
  Component spheres of diameters $d_p = D_s$ are used to represent the disk-shaped particles.  The
  component spheres are arranged in a hexagonal lattice with $n_l = [D_l / d_p]$ spheres in the
  longitudinal direction and $n_i = [D_i/d_p]$ spheres in the transverse direction, where $[\cdot]$
  indicates rounding to the nearest integer. This is illustrated in Fig.~\ref{fig:layout-table}(a).

\item Blade-shaped particle are those whose intermediate axis is much shorter than the long axis
  ($D_i < \frac{2}{3} D_l$) and short axis much shorter than the intermediate axis ($D_s <
  \frac{2}{3} D_i$). The component sphere-based representation of blade-shaped particles is similar
  to that of the disks with lattice size $n_l$ and $n_i$ in the longitudinal and transverse
  directions, respectively, as shown in Fig.~\ref{fig:layout-table}(b).

\item Rod-shaped particles are those whose short and intermediate axes are similar length
  ($\frac{2}{3}D_i < D_s < D_i$) but the long axis is much longer ($D_i < \frac{2}{3}
  D_l$). Component spheres of diameter $d_p = \sqrt{D_s D_i}$ are used to represent these particles,
  with $n_l = [D_l/d_p]$ particles in the longitudinal direction. This is illustrated in
  Fig.~\ref{fig:layout-table}(c). 

\item Equant particles are those whose long, intermediate, and short axes are of similar length, i.e.,
  $\frac{2}{3}D_l < D_i < D_l$ and $\frac{2}{3}D_i < D_s < D_i$.  Note that in the algorithm
  described above for constructing bonded-sphere representation of disk-, blade-, and rod-shaped
  particles, component spheres in the same composite particle have a uniform diameter
  $d_p$. This simplification is adopted to reduce the number of particles and to avoid overly
  complicated shapes. However, restricting component spheres to one size is not practical for
  representing equant particles unless each equant particle is represented with a single particle of
  diameter $d_p = (D_l \, D_i \, D_s)^{1/3}$. However, this simplification would ignore the
  nonspherical characteristics of the particle, which can be significant when $D_i \approx
  \frac{2}{3}D_l$ and/or $D_s \approx \frac{2}{3}D_i$. In the proposed method, we still attempt to
  use the smallest number of component spheres to represent each particle and maximize the wetted
  surface for each component sphere. To this end, the equant particles are further classified to four
  subtypes, each represented with one main sphere and zero to three auxiliary spheres depending on
  the particle geometry. The discription of the representation of each subtype of equant particles
  is detailed in the Appendix.

\end{itemize}

We emphasize that the bonded-sphere representation is a drastic simplification compared to those
employed in dry DEM of industrial processes. We only attempt to preserve the ratios of the three
axes and not the exact shape, since the information on the latter is rarely available for natural
sediments. We will use numerical simulations to show that by preserving the axis aspect ratios, the
mass, moment of inertia, fall velocity, incipient motion, and transport rate can be reproduced.

\subsection{Implementation and Numerical Methods}

The numerical simulations are performed by using the hybrid CFD--DEM solver \textit{SediFoam} which
is a particle-laden flow solver with emphasis on sediment transport. The solver has been validated
by the authors extensively for sediment transport applications~\citep{sun16awr,sun2016sedi}.
\textit{SediFoam} is developed based on two state-of-the-art open-source codes: A CFD platform
OpenFOAM (Open Field Operation and Manipulation) developed by \citet{openfoam} is employed to solve
the fluid flow, and a molecular dynamics simulator LAMMPS (Large-scale Atomic/Molecular Massively
Parallel Simulator) developed at the Sandia National Laboratories~\citep{lammps} is employed to
predict the motion of sediment particles. The interface of OpenFOAM and LAMMPS is implemented for
the communication of shared information in parallel. The code is publicly available at
https://github.com/xiaoh/sediFoam under GPL license.  Detailed introduction of the implementations
are discussed in our previous work~\citep{sun2016sedi}.

The solution algorithm of the fluid solver is partly based on the work of~\cite{rusche03co} on
bubbly two-phase flows. PISO (Pressure Implicit Splitting Operation) algorithm is used to prevent
velocity--pressure decoupling on a co-located grid~\citep{issa86so}.  A second-order implicit scheme
is used for time integrations. A second-order central scheme is used for the spatial discretization
of convection terms and diffusion terms. An averaging algorithm based on diffusion is implemented to
obtain smooth $\varepsilon_s$, $\xmb{U}_s$ and $\xmb{F}^{fp}$ fields from discrete sediment
particles~\citep{sun14db2, sun14db1}. To resolve the collision between the sediment particles, the
contact force between sediment particles is computed with a linear spring-dashpot
model~\citep{cundall79}. The time step to resolve the particle collision is less than 1/50 the
contact time to avoid particle inter-penetration~\citep{sun07ht}. The algorithm in
Fig.~\ref{fig:layout-table} is used to construct irregular particles before numerical simulation.
The translation and rotation of the irregular particles are solved by the rigid body dynamics solver
in LAMMPS~\citep{miller02sqs,ikeguchi04prb}.

\section{A Priori Tests of Properties and Sedimentation Characteristics of Single Particles}
\label{sec:veri}

\begin{figure}[htbp]
  \centering
  \includegraphics[width=0.55\textwidth]{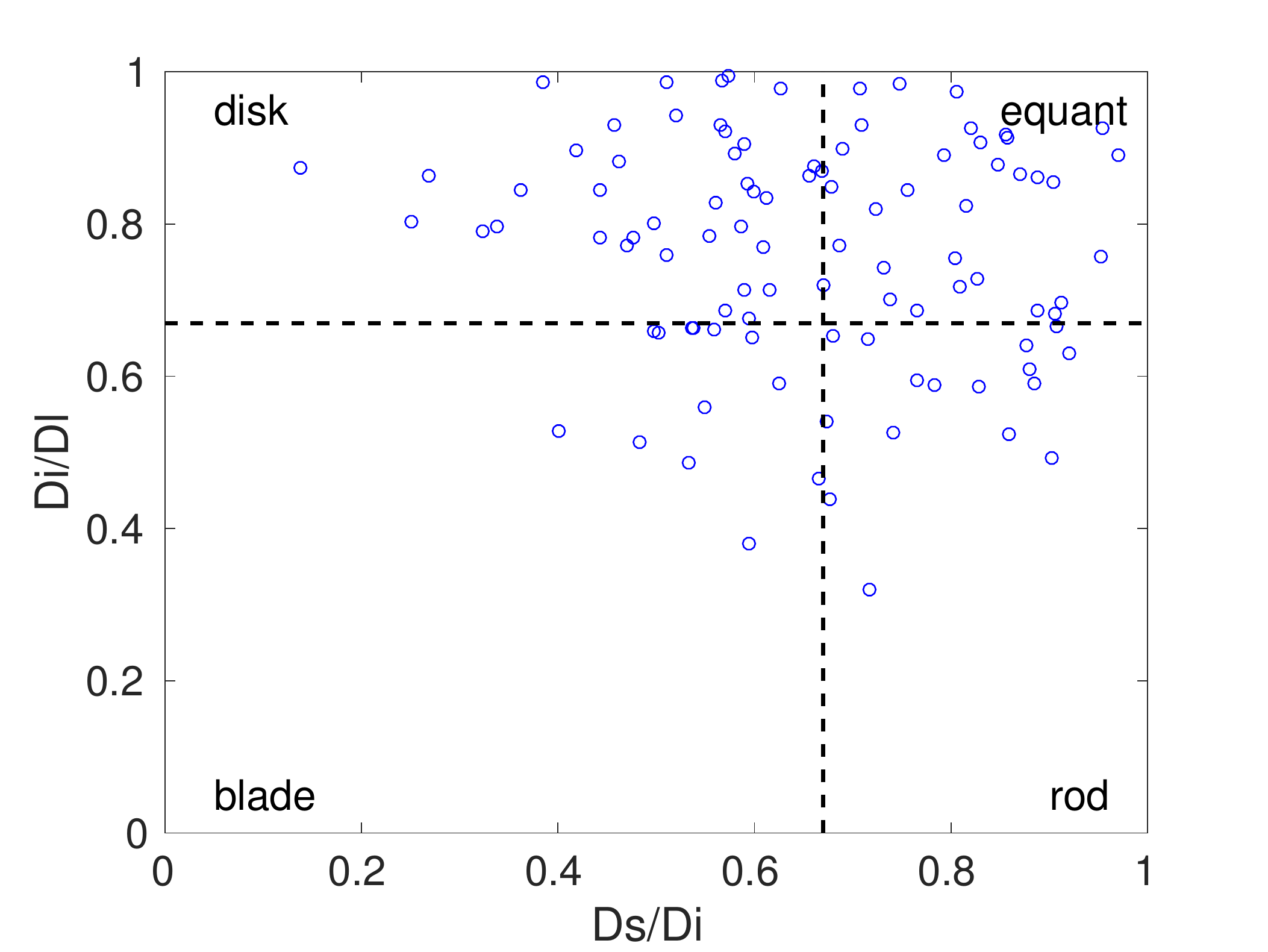}
  \caption{Classification of the shapes of the irregular particles in the a priori test.}
  \label{fig:classification}
\end{figure}

\begin{figure}[htbp]
  \centering
  \subfloat[mass]{
  \includegraphics[natheight = 500, natwidth = 700,width=0.45\textwidth]{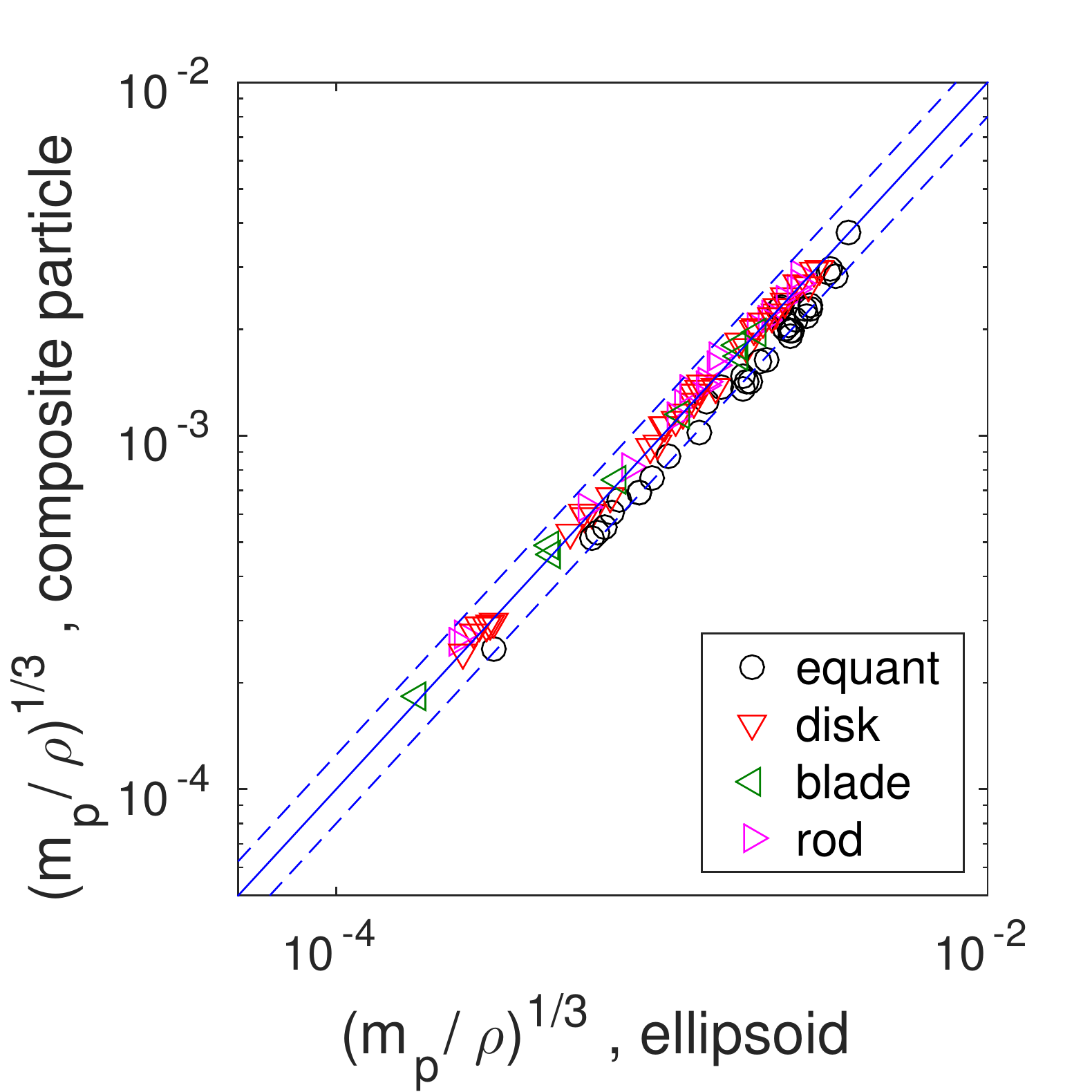}
  }
  \subfloat[moment of inertia]{
  \includegraphics[natheight = 500, natwidth = 700,width=0.45\textwidth]{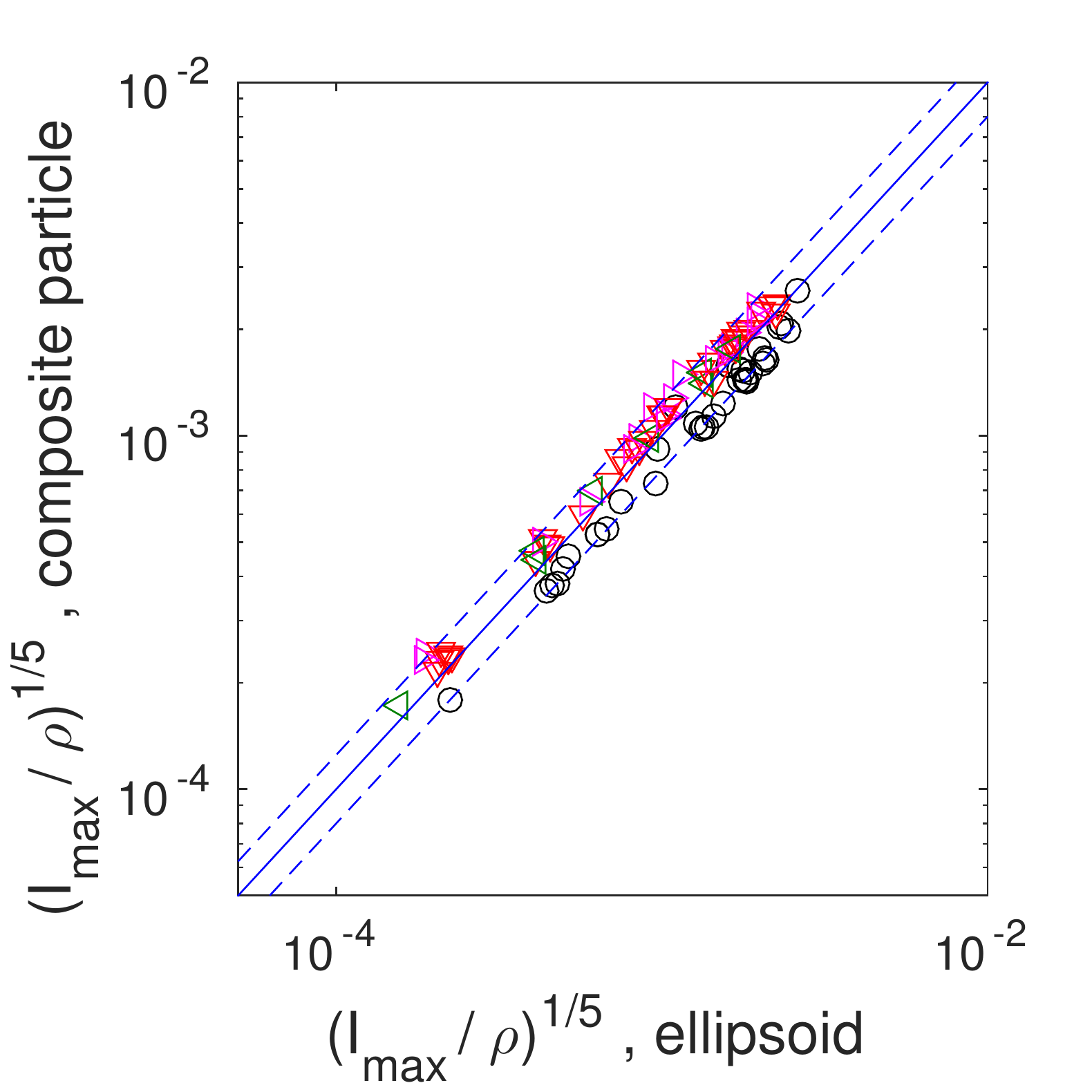}
  }
  \caption{The mass and moment of inertia of composite irregular particles compared with
  ellipsoidal particles.}
  \label{fig:compare-ellipsoid}
\end{figure}

\begin{figure}[htbp]
  \centering
  \subfloat[mass]{
  \includegraphics[natheight = 500, natwidth = 700,width=0.45\textwidth]{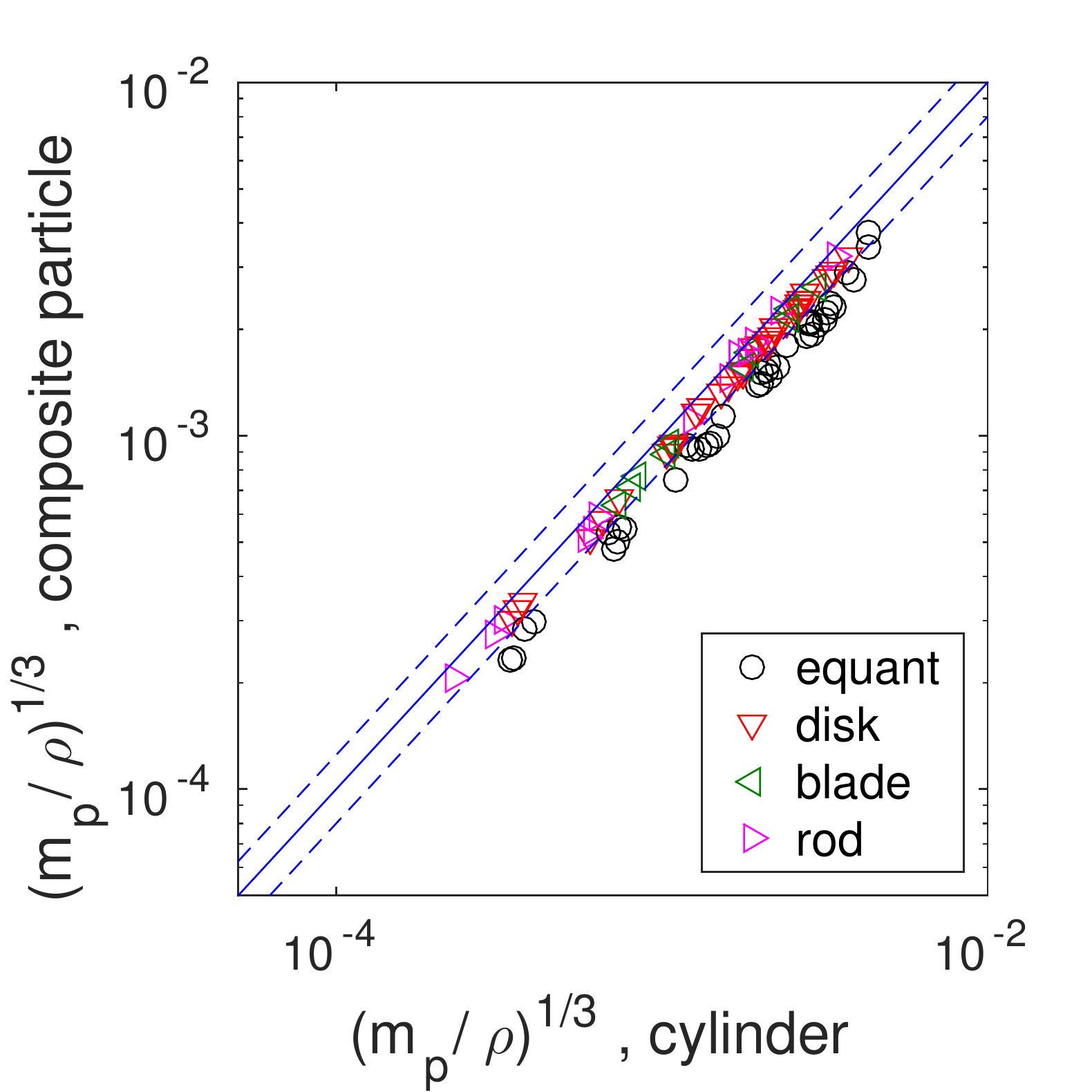}
  }
  \subfloat[moment of inertia]{
  \includegraphics[natheight = 500, natwidth = 700,width=0.45\textwidth]{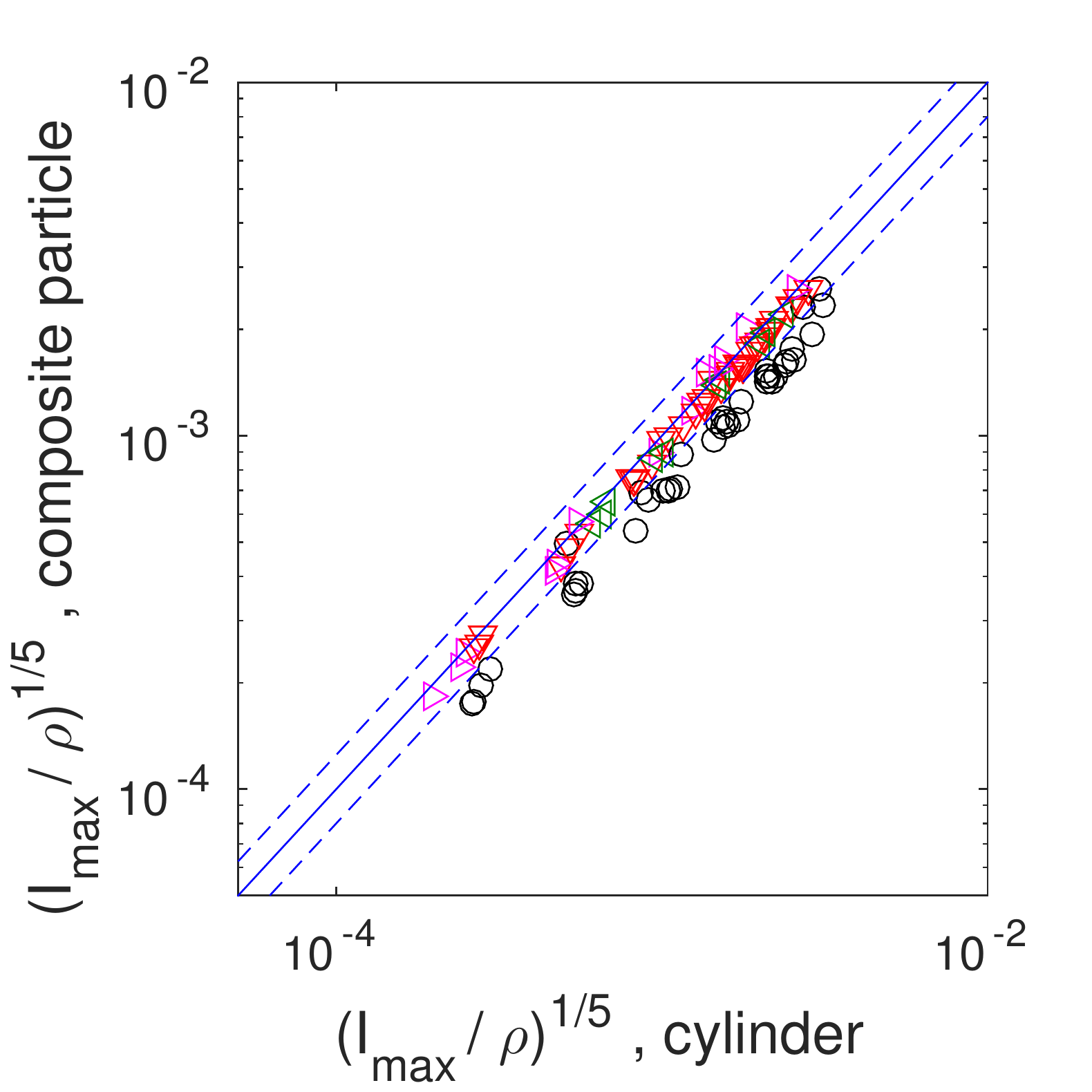}
  }
  \caption{The mass and moment of inertia of composite irregular particles compared with
  cylindrical particles.}
  \label{fig:compare-cylinder}
\end{figure}

The first objective of the a priori test is to show that the proposed algorithm lead to particles
with approximately the same mass and moment of inertia as the irregular particle that we aim to
represent. However, since only the lengths $D_l$, $D_i$, and $D_s$ of the three axes are used to
construct the bonded-sphere representation, the exact mass and moment of inertia of irregular
particle are unknown. Therefore, we generated 100 constructed particles over a wide range of values
of $D_l$, $D_i$, and $D_s$, and compared them with two representative particle shapes: (1) an
ellipsoid of three axes of length $D_l$, $D_i$, and $D_s$, respectively, and (2) a cylinder of
length $D_l$ whose cross-section is an ellipse of axis $D_i$ and $D_s$. Both $D_s/D_i$ and $D_l/D_i$
of the constructed particles are selected randomly. It can be seen in Fig.~\ref{fig:classification}
that the proportion of the each type of irregular particle in the a priori test is consistent with
the measurement by~\cite{smith03ef}. The long axes $D_l$ of the irregular particles range from
0.2~mm to 2~mm according to size of calcareous Sand in experimental measurement~\citep{smith03sc}.
The comparison between the constructed particles and the ellipsoidal reference particles with the
same axes in mass and moment of inertia are shown in Fig.~\ref{fig:compare-ellipsoid}. Each point in
the scatter plot represents an irregular particle with a different combination of $D_l$, $D_i$, and
$D_s$ values. The solid line plotted in the figure indicates perfect agreement; two dash lines of
slope 0.80 and 1.25 indicate under-prediction and over-prediction by 25\%, respectively. It should
be noted that the two dashed lines are parallel with the solid line when plotted using logarithmic
scale. It can be seen in Fig.~\ref{fig:compare-ellipsoid} that the difference in mass and moment of
inertia between the constructed particles and the ellipsoidal reference particles is within 25\%.
The comparisons of the mass and moment of inertia between the reconstructed particle and the
cylindrical reference particles are shown in Fig.~\ref{fig:compare-cylinder}.  It is evident from
the figure that the mass and the moment of inertia of the cylindrical reference particles are larger
than those of the constructed particles.  This is because a cylindrical reference particle has the
maximum mass and moment of inertia of a particle for given $D_l$, $D_i$, and $D_s$ values. The
over-prediction of the moment of inertia for some rod-shaped particles shown in
Fig.~\ref{fig:compare-cylinder}(b) is due to the rounding of $D_l/D_i$ to the nearest integer. In
summary, the mass and moment of inertia of the constructed irregular particles are in good agreement
with the irregular particles that we aim to represent.

Another objective of the a priori test is to examine the sedimentation characteristics of the
composite particles. To this end, the constructed particles of two representative irregular
particles are released in quiescent water at various random initial angles, and their trajectory and
change of orientation are observed. It is found that regardless of the initial orientation of the
particles when released, they adjust to an orientation where the maximum projection is normal to the
fall velocity direction. This is consistent with the experimental observations reported in the
literature~\citep{tran2004drag}. However, previous authors have also reported the spiral motions and
the randomness of particle trajectory~\citep{tran2004drag}, which are not observed in our
simulations. This is attributed to the fact that the instantaneous shear instability and vortex
shedding that lead to spiral motion is not represented in CFD--DEM. Specifically, since the
fluid--particle interface is not explicitly resolved in CFD--DEM, only the mean effect of the
particle on the flow is represented by applying forces at the cell centers surrounding the
particle. However, as with many other dense-phase particle-laden flows, the dynamics of sediment
transport is dominated by a number of competing mechanisms such as particle collisions and the
associated granular flow dynamics, boundary layer turbulence, and fluid--particle
interactions. Therefore, it is not essential or feasible to represent each mechanism exactly.  In
fact, this compromise is what allows CFD--DEM to handle a much larger number of particle than
interface-resolved methods ~\citep{kempe14ot}.

\begin{figure}[htbp]
  \centering
  \includegraphics[width=0.45\textwidth]{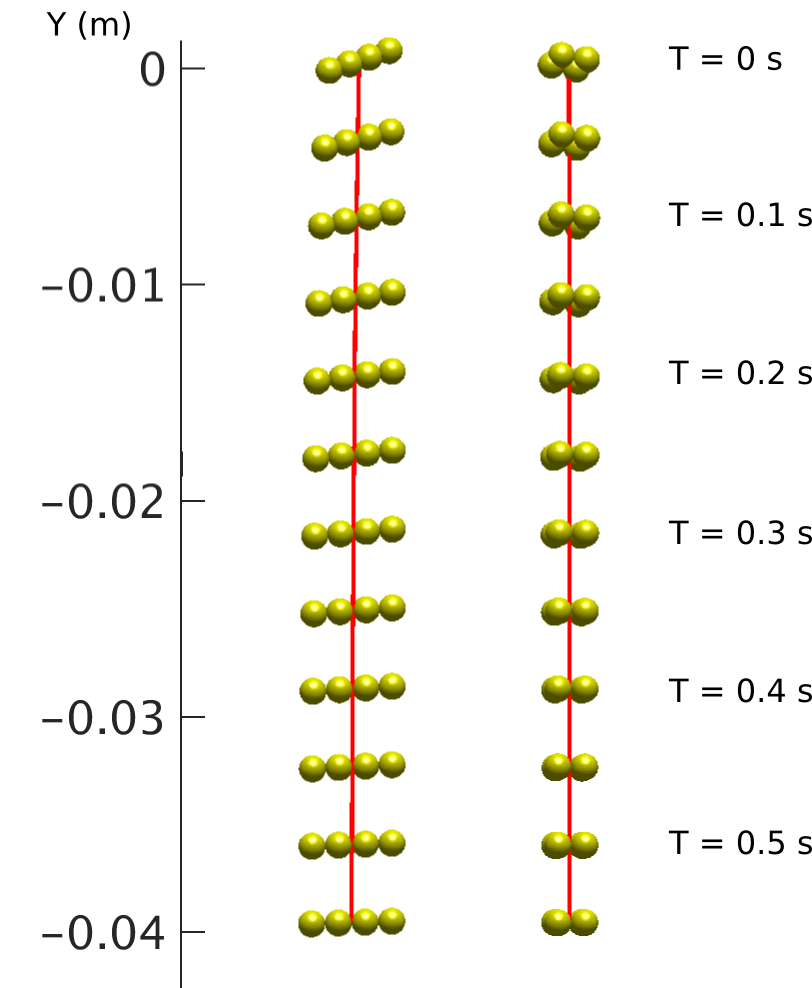}
  \caption{The orientation and trajectory of rod- and disk-shaped particles in sedimentation. The
  red lines are the traces of the center of the composite particle. The particles are enlarged for
  visualization.}
  \label{fig:fall-orientation}
\end{figure}


\section{Test in Sediment Transport Simulations}
\label{sec:vali}
Four numerical tests are performed to demonstrate that the proposed method for representing
irregular shaped particles is capable of predicting integral quantities in sediment transport
applications. The applications include the sedimentation of a single particle in quiescent fluid,
the rotation of the irregular particle in boundary layer, the incipient motion of particles, and the
sediment transport in a periodic channel. The setup of the numerical tests are based on previous
experimental studies~\citep{smith03ef,smith04im,smith05tr}.

\begin{figure}[htbp]
  \centering
  \subfloat[sedimentation test]{
  \includegraphics[width=0.3\textwidth]{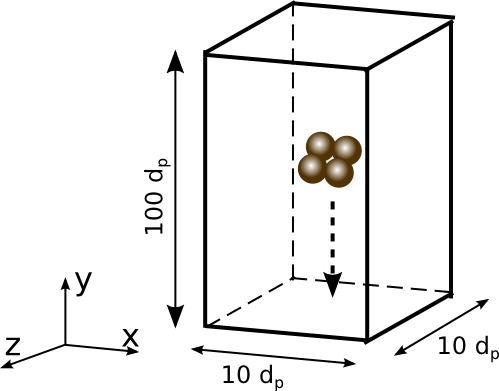}
  }
  \subfloat[rotation test]{
  \includegraphics[width=0.55\textwidth]{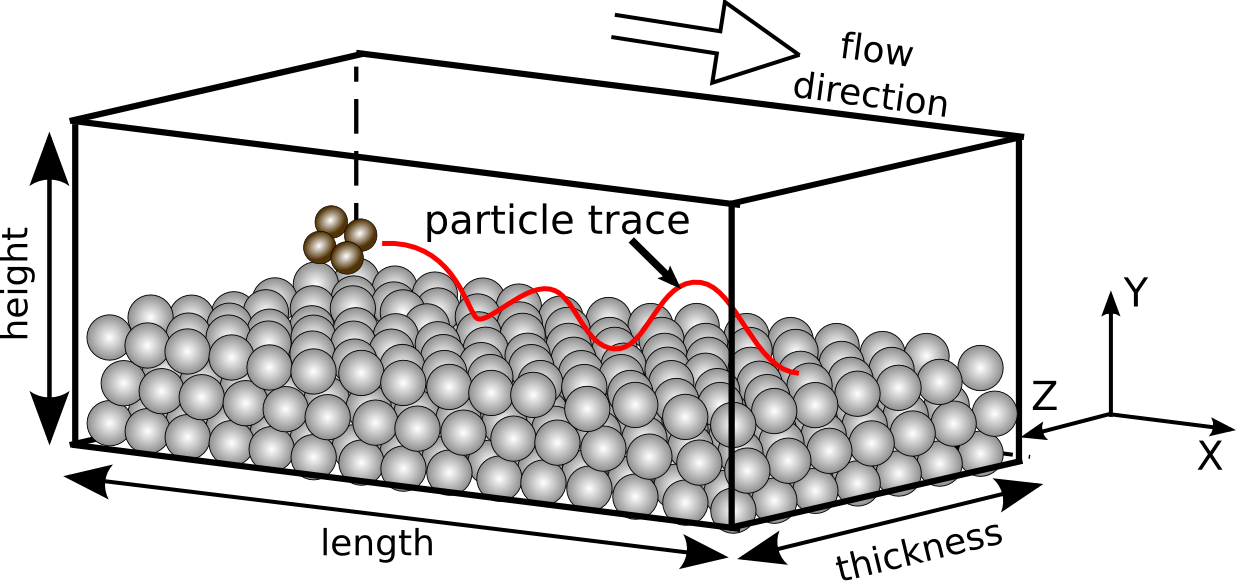}
  }
  \vspace{0.1in}
  \subfloat[incipient motion test and sediment transport test]{
  \includegraphics[width=0.55\textwidth]{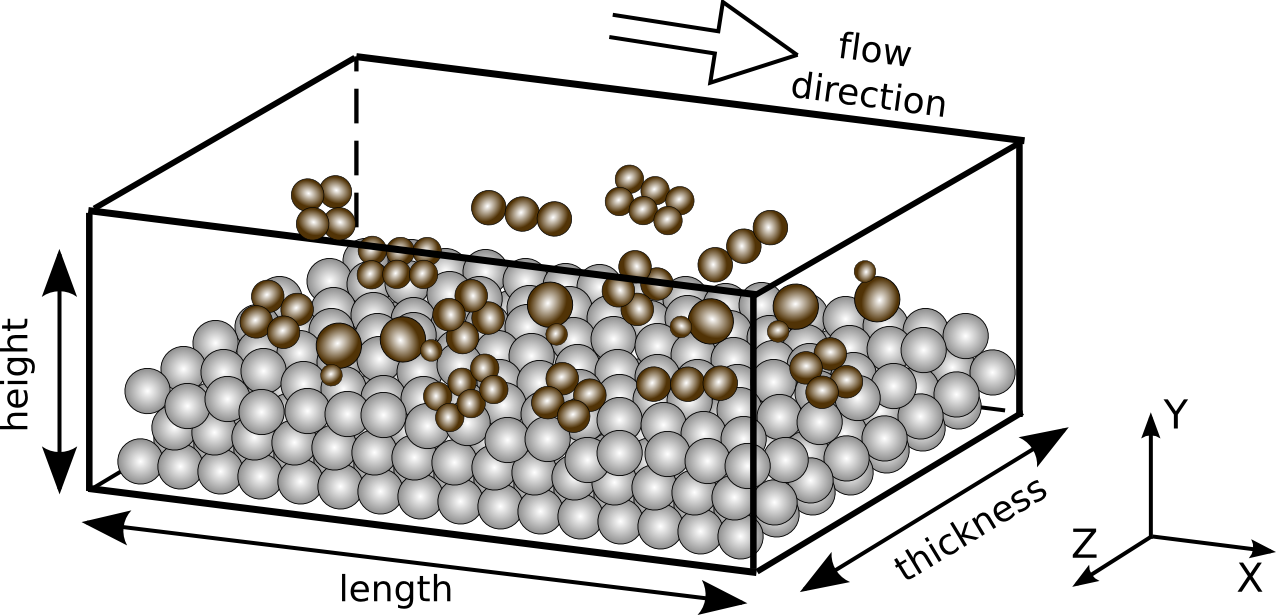}
  }
  \caption{Layout of the numerical simulations of sediment transport cases: (a) sedimentation test
  of the terminal velocity, (b) rotation test in boundary layer test, and (c) incipient motion test
  and sediment transport test. The brown particles are moving irregular particles; the gray
  particles are fixed particles to provide rough wall boundary conditions. Note that the first case
  in rotation test are performed without using the rough bottom wall but are not plotted. The
  particle sizes are not to scale.}
  \label{fig:layout-all}
\end{figure}

The geometry of the simulation is shown in Fig.~\ref{fig:layout-all}.  The dimensions of the domain,
the mesh resolutions, and the fluid and particle properties used are detailed in
Table~\ref{tab:param-all}. Periodic boundary condition is applied in both $x$- and $z$-directions.
For the pressure field, zero-gradient boundary condition is applied in $y$-direction; for the
velocity field, no-slip wall condition is applied at the bottom in $y$-direction, and slip wall
condition is applied on the top. In the sedimentation test, the initial flow is quiescent, and a
uniform grid mesh is used.  In the rotation test, the fluid flow is driven by a pressure gradient to
maintain a constant flow rate $q_f$. In addition, different mesh resolutions and boundary conditions
are used in the two cases of the rotation test. In the first rotation case, simulations are
performed using two meshes: (1) a coarse uniform grid mesh, and (2) a fine mesh refined in the
vertical ($y$-) direction towards its bottom boundary. No fixed particles are used at the bottom
wall. In the second rotation case, the CFD mesh is refined at the bottom, and three layers of fixed
particles are arranged hexagonally to provide a rough bottom boundary condition to the moving
particles, as is shown in Fig.~\ref{fig:layout-all}(b). In the incipient motion test and sediment
transport test, the CFD mesh is refined at the bottom, the flow rate is driven by a pressure
gradient, and the rough bottom wall is applied. Since detailed size distribution of the irregular
particle is not known, the size of irregular particles used in the two tests is uniform.  Four
example types of bonded particles, shown in Fig.~\ref{fig:layout-all}(c), represent each type of
natural sediment particle, and the proportion of each type is consistent with the measurement
by~\cite{smith03ef}.

\begin{table}[!htbp]
  \caption{Parameters of the numerical simulations. The nominal diameter $D_n = (D_lD_iD_s)^{1/3}$
is used to normalize the dimensions of the computational domain. }
 \begin{center}
 \begin{tabular}{lcccc}
   \hline       & \specialcell{sedimentation \\ test}
                & \specialcell{rotation \\ test}
                & \specialcell{incipient \\ motion test}
                & \specialcell{sediment \\ transport test} \\
   \hline
   domain dimensions                            &\\
   \qquad width $(L_x/D_n$)                     & 10   & 24  & 144 & 144 \\
   \qquad height $(L_y/D_n$)                    & 100  & 20  & 160  & 160 \\
   \qquad transverse thickness $(L_z/D_n$)      & 10   & 12  & 72  & 72 \\
   mesh resolutions                             &\\                        
   \qquad width $(N_x)$                         & 10    & 12    & 120  & 120 \\
   \qquad height $(N_y)$                        & 100   & 12 or 30    & 100  & 100 \\
   \qquad transverse thickness $(N_z)$          & 10    &  6    & 60  & 60 \\
   particle properties & \\
   \qquad total number of irregular particle    & 1     & 1   & $6.4\times10^4$ & $6.4\times10^4$  \\
   \qquad nominal diameter $D_n$~[mm]                   & 0.7--5.5   & 0.5   & 0.2--0.8 & 0.2--0.8\\
   \qquad density~$\rho_s$~[$\times 10^3~\mathrm{kg/m^3}$]  & \multicolumn{4}{ c }{2650} \\
   \qquad particle stiffness coefficient~[N/m]  & \multicolumn{4}{ c }{20} \\
   \qquad normal restitution coefficient        & \multicolumn{4}{ c }{0.01} \\
   \qquad coefficient of friction               & \multicolumn{4}{ c }{0.4} \\
   fluid properties & \\
   \qquad viscosity $\nu$~[$\times 10^{-6}~\mathrm{m^2/s}$]  & 1.0     & 10   & 1.0 & 1.0  \\
   \qquad density $\rho_f$~[$\times 10^3~\mathrm{kg/m^3}$] & \multicolumn{4}{ c }{1.0} \\
   \hline
  \end{tabular}
 \end{center}
 \label{tab:param-all}
\end{table}

\subsection{Sedimentation of a Single Particle in Quiescent Flow}
\begin{figure}[htbp]
  \centering
  \subfloat[terminal velocity]{
  \includegraphics[natheight = 500, natwidth = 700,width=0.495\textwidth]{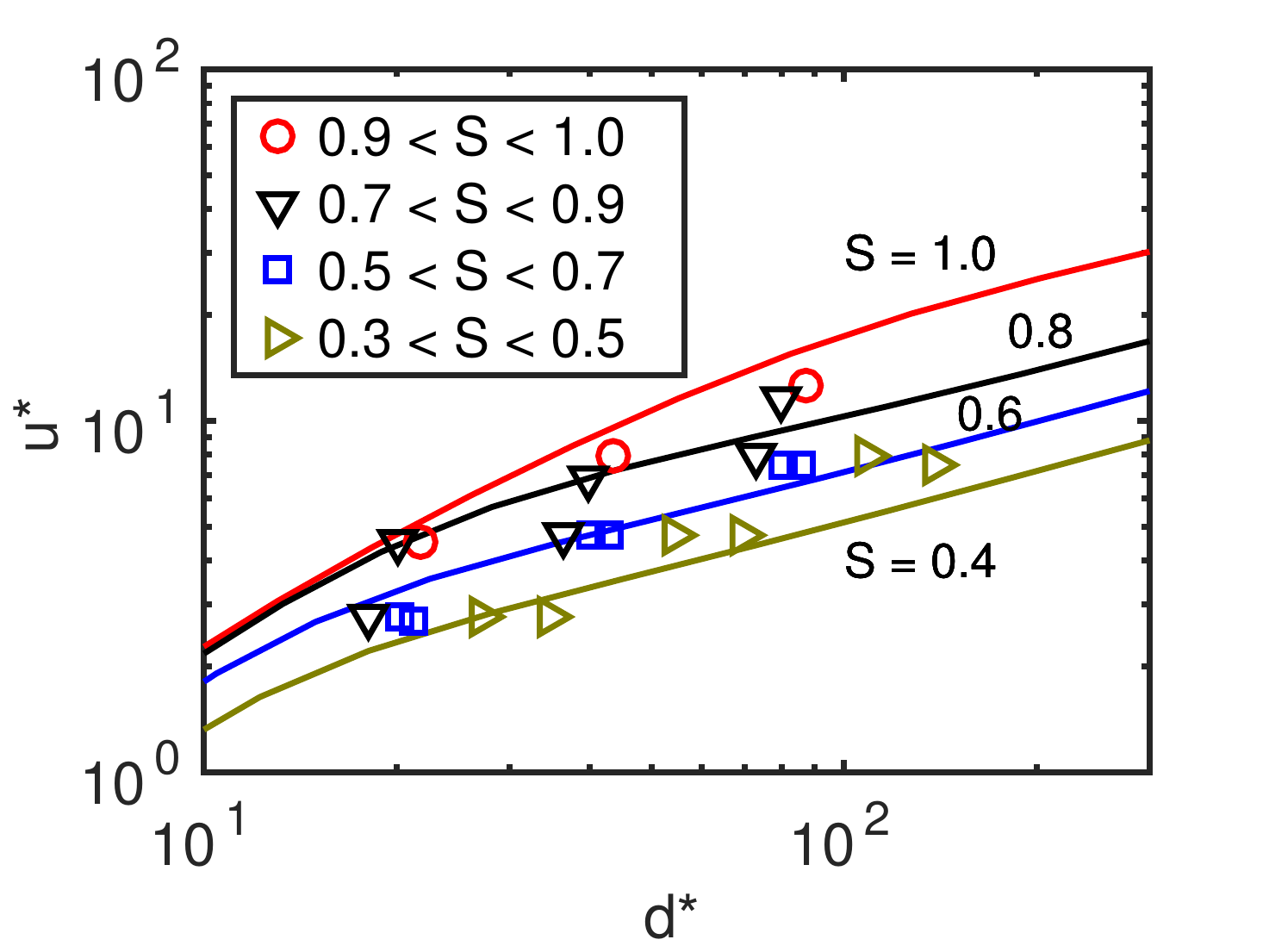}
  }
  \subfloat[drag coefficient]{
  \includegraphics[natheight = 500, natwidth = 700,width=0.495\textwidth]{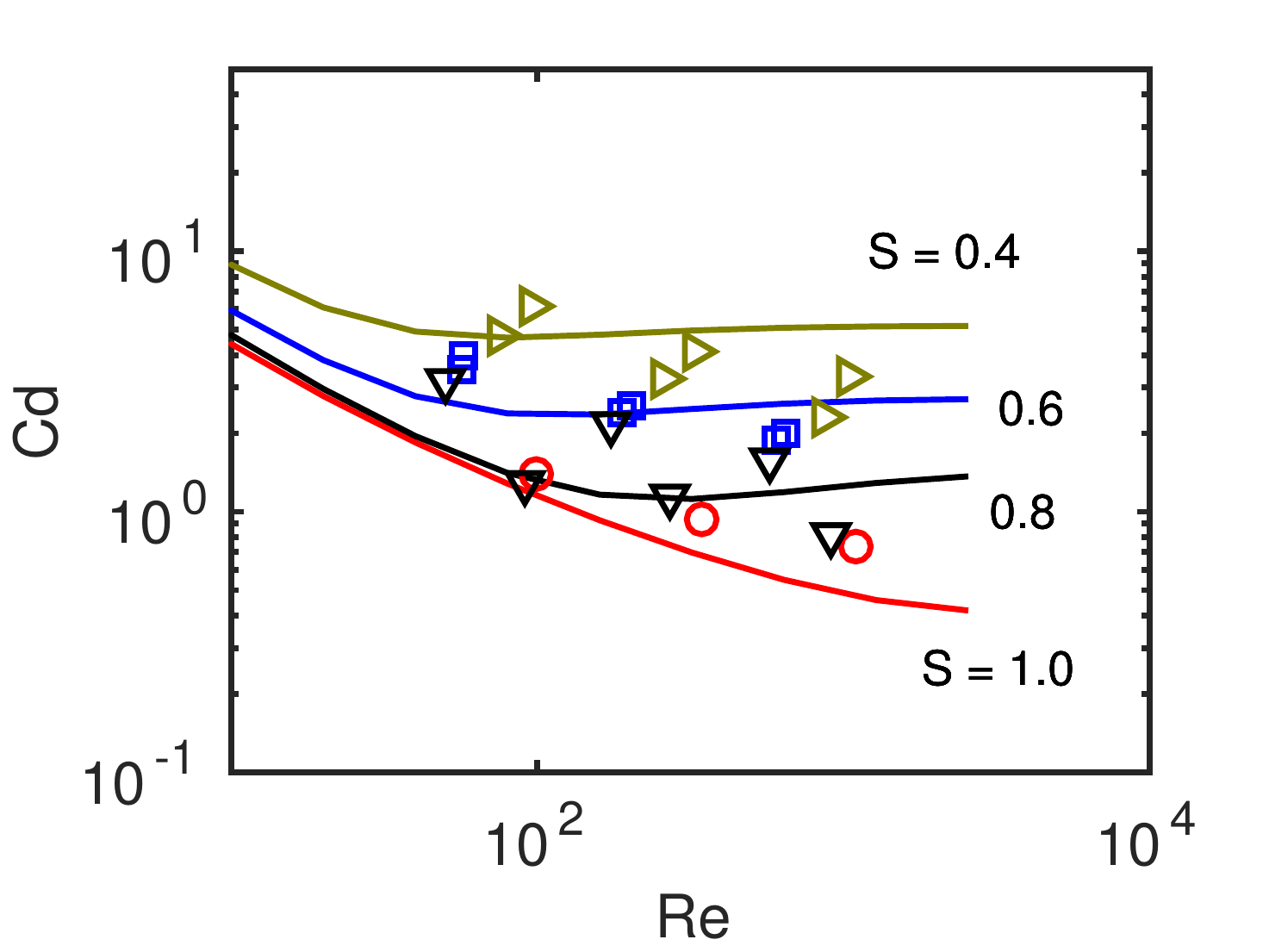}
  }
  \caption{The terminal velocity and drag coefficient of composite irregular particles of various
    sizes and shapes: (a) non-dimensional terminal velocity $u^*$ plotted as a function of
    non-dimensional particle diameter $d^*$; (b) drag coefficient $C_d$ plotted as a function of
    Reynolds number.}
  \label{fig:termU}
\end{figure}

The sedimentation test of the constructed particles is performed to show the terminal velocity and
drag coefficient of the constructed particles are consistent with the experimental results. The
simulations of the sedimentation of 21 irregular particles of various shapes are performed. The
nominal diameter $D_n = (D_lD_iD_s)^{1/3}$ of the constructed particles ranges from 0.7~mm to 5.5~mm
according to the experiments conducted by~\cite{smith03sc}. Figure~\ref{fig:termU} shows the
terminal velocity and drag coefficient of constructed particles of different shape factors $S$. The
regression curves proposed by~\cite{haider89dc} are used to compare with the results. The shape
factor is defined as $S = S_{irregular}/S_{sph}$, where $S_{irregular}$ refers to the surface area
of the irregular particle; $S_{sph}$ is the surface area of a spherical particle with the same
volume as the irregular particle.  Non-dimensional terminal velocity $u^*$ is defined as:
\begin{equation}
  u^* = u_t\left(\frac{\rho_f^2}{g\mu(\rho_s-\rho_f)}\right)^{1/3},
  \label{eq:uterm-dimless}
\end{equation}
where $u_t$ is the terminal velocity of the sediment particles. The non-dimensional particle
diameter $d^*$ is defined as:
\begin{equation}
  d^* = D_{n}\left(\frac{g\rho_f(\rho_s-\rho_f)}{\mu^2}\right)^{1/3},
  \label{eq:dp-dimless}
\end{equation}
where the nominal diameter $D_{n}$ is the diameter of the equivalent sphere particle having the same
volume. It can be seen in Fig.~\ref{fig:termU} that the agreement of the terminal velocities and
drag coefficients between the numerical simulations and the experimental measurements is good for
particles at different shape factors. This indicates the proposed approach of representing irregular
particles can capture the decrease of terminal velocity and the increase of drag coefficient
compared with spherical particles. In addition, Fig.~\ref{fig:termU} shows that the scattering of
the terminal velocity and drag coefficient is non-negligible for sediment particles at similar size
and shape factor. This is because the shape factor is inadequate to describe the settling
characteristics of different types of irregular particles. For example, the terminal velocity of
rod- and disk-shaped particles of same shape factor can differ significantly. This scattering in
terminal velocity and drag coefficient of irregular particles is also observed in previous
experimental measurements~\citep{haider89dc,smith03sc}. 

\subsection{Translation and Rotation of Particles in Boundary Layer}

The motion of irregular particle in the boundary layer is investigated to demonstrate that the
proposed approach to represent irregular particle can capture the rotation of sediment particles due
to irregularity. Two numerical tests are performed to (1) show the flow velocity gradient is
critical to the rotation of irregular particle, and to (2) compare the rotations of different types
of irregular particles. In addition, the factors that can influence irregular particle rotation are
investigated quantitatively. In the tests, the flow in the periodic channel is laminar at $Re =
2000$.

\begin{figure}[htbp]
  \centering
  \subfloat[fine mesh]{
  \includegraphics[width=0.85\textwidth]{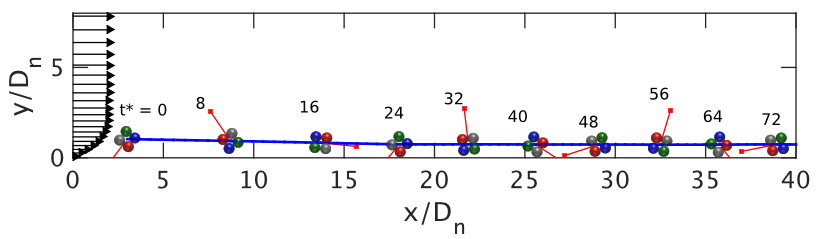}
  }
  \hspace{0.1in}
  \subfloat[coarse mesh]{
  \includegraphics[width=0.85\textwidth]{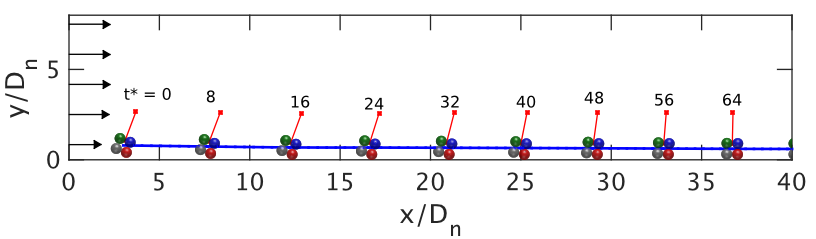}
  }
  \hspace{0.1in}
  \subfloat[zoom-in view of fine mesh]{
  \includegraphics[width=0.45\textwidth]{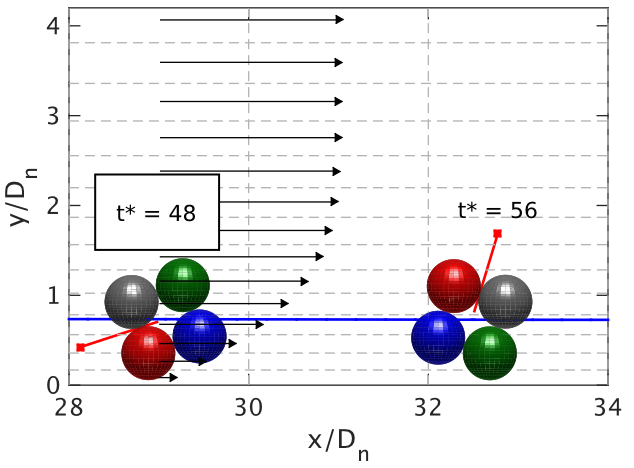}
  }
  \subfloat[zoom-in view of coarse mesh]{
  \includegraphics[width=0.45\textwidth]{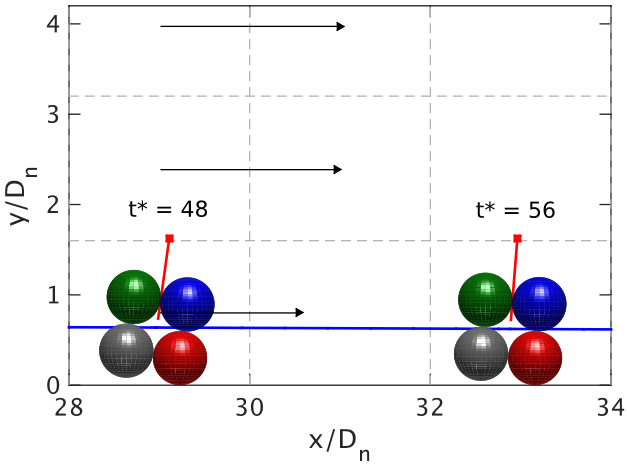}
  }
  \caption{The motion of a disk-shaped particle on a smooth bed obtained by using different meshes
  in laminar flow: (a) fine mesh; (b) coarse mesh; (c) zoom-in view of fine mesh; (d) zoom-in view
  of coarse mesh. The non-dimensional time is defined as $t^* = (t-t_0)U_b/D_n$, where $t_0$ is the
  time to start plotting. The arrows indicates the magnitude and direction of the fluid flow. The red
  flags illustrate the rotation of the disk-shaped particle. The CFD mesh used in the simulations are
  represented using dashed lines.}
  \label{fig:rotation-smooth}
\end{figure}

To demonstrate the influence of flow velocity gradient is significant to the rotation of irregular
particle, simulations on different mesh resolutions are performed. In the fine mesh test, the flow
velocity gradient at the length scale of the composite particle diameter $D_n$ is resolved, and the
CFD mesh is refined in the vertical direction at the bottom.  In the coarse mesh test, the cell size
is comparable to the composite particle diameter $D_n$, and thus the velocity gradient of the flow
at the scale of the particle diameter is not resolved. The predictions of the motion of a
disk-shaped particle on meshes of different resolutions are shown in Fig.~\ref{fig:rotation-smooth}.
It can be seen in Fig.~\ref{fig:rotation-smooth}(a) that the rotation of disk-shaped particle is
captured when using fine mesh; whereas Fig.~\ref{fig:rotation-smooth}(b) shows that the particle is
sliding on the bed when coarse mesh is used. The difference in the motion of sediment particles is
due to difference of the flow velocity variation $\Delta u = D_n \frac{\partial U_f}{\partial x}$.
When the gradient of flow velocity is resolved using fine mesh, shown in
Fig.~\ref{fig:rotation-smooth}(c), the drag force on the top part of the irregular particle is large
and rotates the particle about its center of mass.  In contrast, when using coarse mesh, although
the flow velocity is large enough to move the sediment particle via the drag forces, the torque
excerted on the particle by the flow, as represented by the coarse mesh, is not enough to cause
appreciable rotates. This test demonstrates that the proposed method of representing irregular
particles can capture the dominant mechanism casuing the particle rotation in the boundary layer.

\begin{figure}[htbp]
  \centering
  \subfloat[disk-shaped particle]{
  \includegraphics[width=0.8\textwidth]{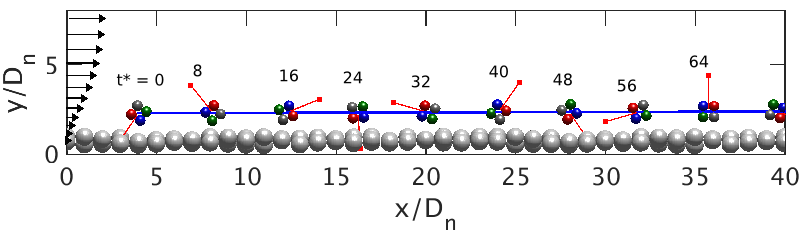}
  }
  \hspace{0.02in}
  \subfloat[blade-shaped particle]{
  \includegraphics[width=0.8\textwidth]{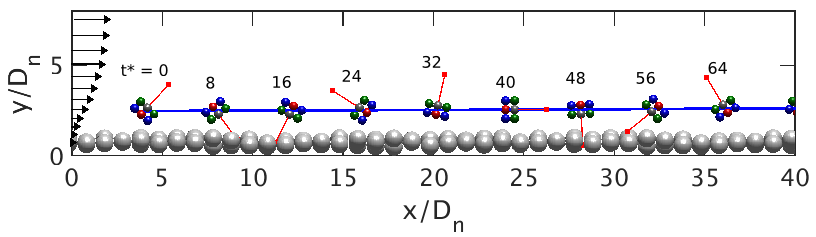}
  }
  \hspace{0.02in}
  \subfloat[rod-shaped particle]{
  \includegraphics[width=0.8\textwidth]{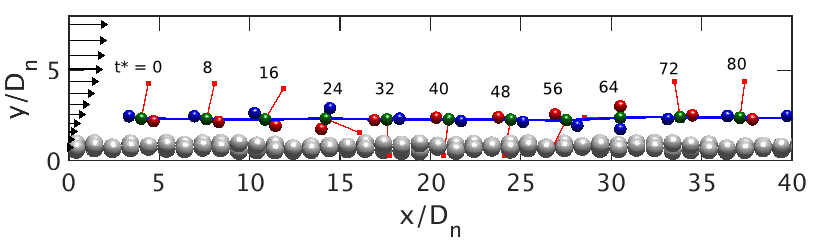}
  }
  \hspace{0.02in}
  \subfloat[equant particle]{
  \includegraphics[width=0.8\textwidth]{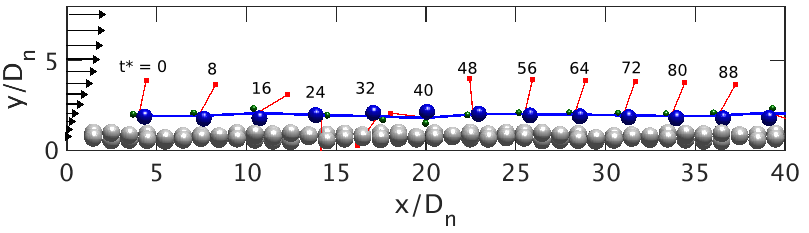}
  }
  \caption{The motion of the irregular particle on the rough bottom: (a) composite disk-shaped
    particle; (b) composite rod-shaped particle; (c) composite blade-shaped particle; (d) composite
    equant particle. The non-dimensional time is defined as $t^* = (t-t_0)U_b/D_n$, where $t_0$ is
    the start time to plot. The rought bottom consists of three layers of fixed particles, but only
    the top layer is plotted.  }
  \label{fig:rotation-rough}
\end{figure}

In addition, numerical tests are performed on all four types of irregular particles using fine mesh.
The purpose of this test is to demonstrate the influence of the particle type to its rotation. The
rough wall boundary condition composed of three layers of fixed particles is applied, which is to
show the proposed approach can also capture the particle rotation on a rough wall. In this test, the
nominal diameters $D_n$ of four irregular particles are the same. The snapshots of the rotation of
irregular particles are shown in Fig.~\ref{fig:rotation-rough}. It can be seen in the figure that
the rotation of different types of irregular particles relative to their centers of mass can be
captured. From the comparison between Fig.~\ref{fig:rotation-rough}(a), (b) and (c) that the
rotation velocity of the rod-shaped particle is smaller than that of the disk- and blade-shaped
particles.  This is because the moment of inertia of rod-shaped particle is relatively larger and
thus less likely to rotate compared with disk- and blade-shaped particle.  Moreover, it is shown in
Fig.~\ref{fig:rotation-rough}(d) that the rotation velocity of the equant particle is very small at
non-dimensional time $t^* = (t-t_0)U_b/D_n \in [48,88]$, where $t_0$ it the time to start plotting.
This can be explained by the fact that the irregularity of the equant particle is relatively small,
and thus the torque obtained in boundary flow to drive the rotation of equant particle is also
small. Therefore, the rotation velocity of the equant particle is smaller than those of other types
with larger irregularity. In addition, it can be seen in Fig.~\ref{fig:rotation-rough} that the
velocities of the streamwise transition of different types of particles are not the same. The equant
particle, which has the largest shape factor, moves more slowly than other types of irregular
particles. This is attributed to the fact that  when the irregularity (shape factor $S$) increases,
the terminal velocity of the irregular particle increases and the particle is likely to move slower
in fluid flow.

\begin{figure}[htbp]
  \centering
  \includegraphics[width=0.70\textwidth]{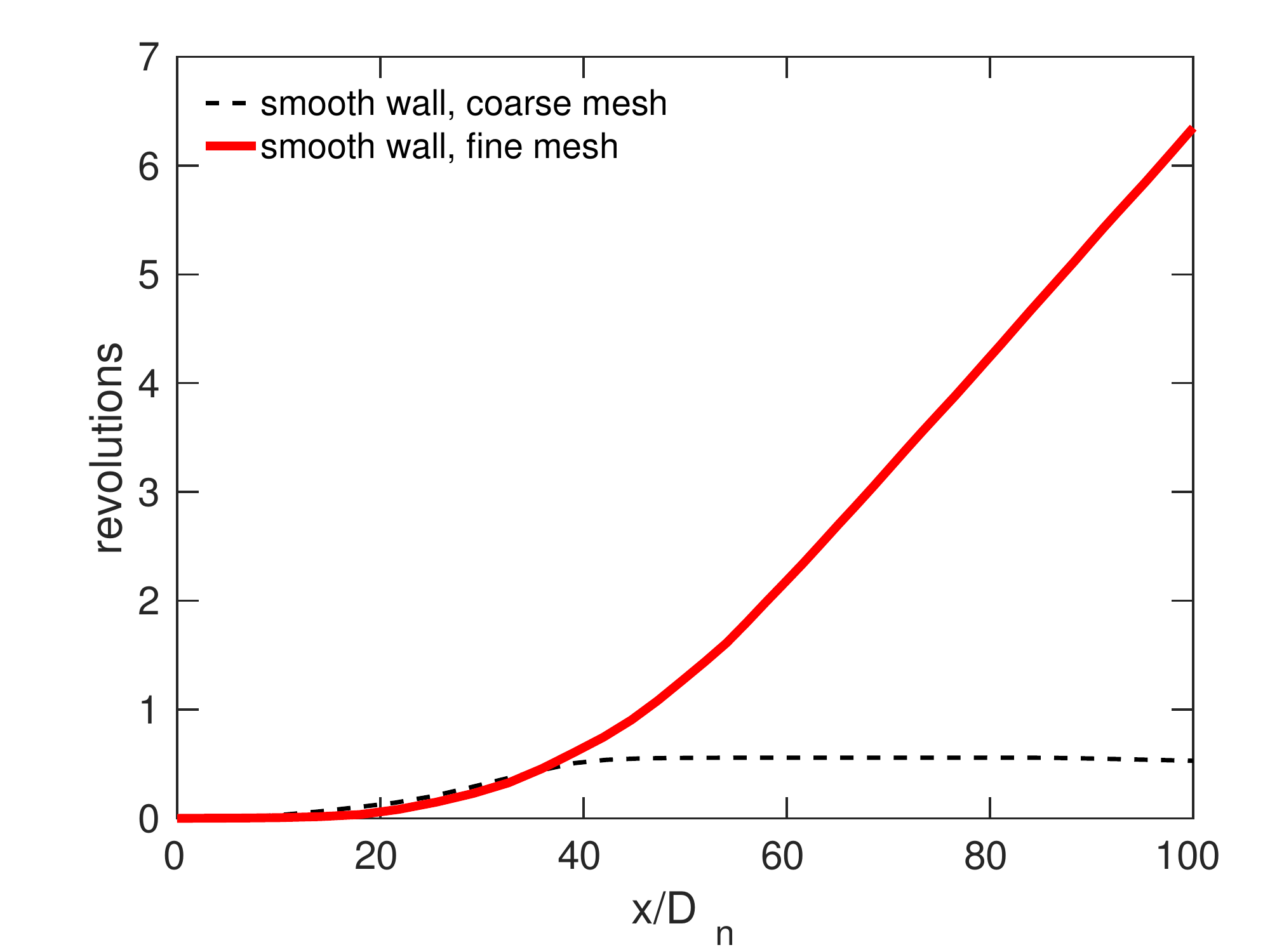}
  \caption{Angle of rotation of the irregular disk-shaped particle in the boundary layer plotted as a
  function of the non-dimensional transitional displacement in the streamwise direction.}
  \label{fig:rotation-angle}
\end{figure}

The numbers of revolutions of the disk-shaped particle plotted as a function of the transitional
displacement in streamwise direction are shown in Fig.~\ref{fig:rotation-angle}. This aims to
investigate the contribution to the particle angular velocity of different factors. The results
shown in the figure are obtained in the previous tests: (1) smooth wall and coarse mesh, (2) smooth
wall and fine mesh, and (3) rough wall and fine mesh. In all three cases, the particle rotates
slowly before it hits the bottom. However, when the disk-shaped particle hits the bottom at $x/D_n =
40$, the angular velocities of the particles increase significantly. This is because the velocities
gradient to drive the particle rotation at the bottom is larger than that in the center of the
channel. It can be also seen in the figure that the angular velocity of particle obtained by using
fine mesh is much larger than that using coarse mesh.  This is attributed to the fact that the
contribution of the flow velocity gradient is captured. In addition, the angular velocity of the
particle obtained using fine mesh on a rough wall is smaller than that on a smooth wall. This is
because the rough wall boundary condition provides more friction than the smooth wall, and thus the
angular velocity of the particle is smaller.

\subsection{Incipient Motion}
\label{sec:incipient}

The incipient motion test of the constructed particles is performed to demonstrate that proposed
approach to represent irregular particles can capture the initiation of motion of them at critical
Shields stress.  Three nominal diameters of the irregular particles used in the test are 0.2~mm,
0.5~mm and 0.8~mm, which is according to the particle diameters selected in~\cite{smith04im}.  For
each particle diameter, the sediment transport rates obtained at five different velocities (ranging
from 0.2~m/s to 0.4~m/s) are plotted as a function of the Shields parameter. To determine the
critical Shields parameter in the incipient motion of irregular particles, the threshold transport
rate is take as $q* = 10^{-4}$ according to~\cite{smith04im}.

\begin{figure}[htbp]
  \centering
  \subfloat[sediment transport rate]{
  \includegraphics[width=0.495\textwidth]{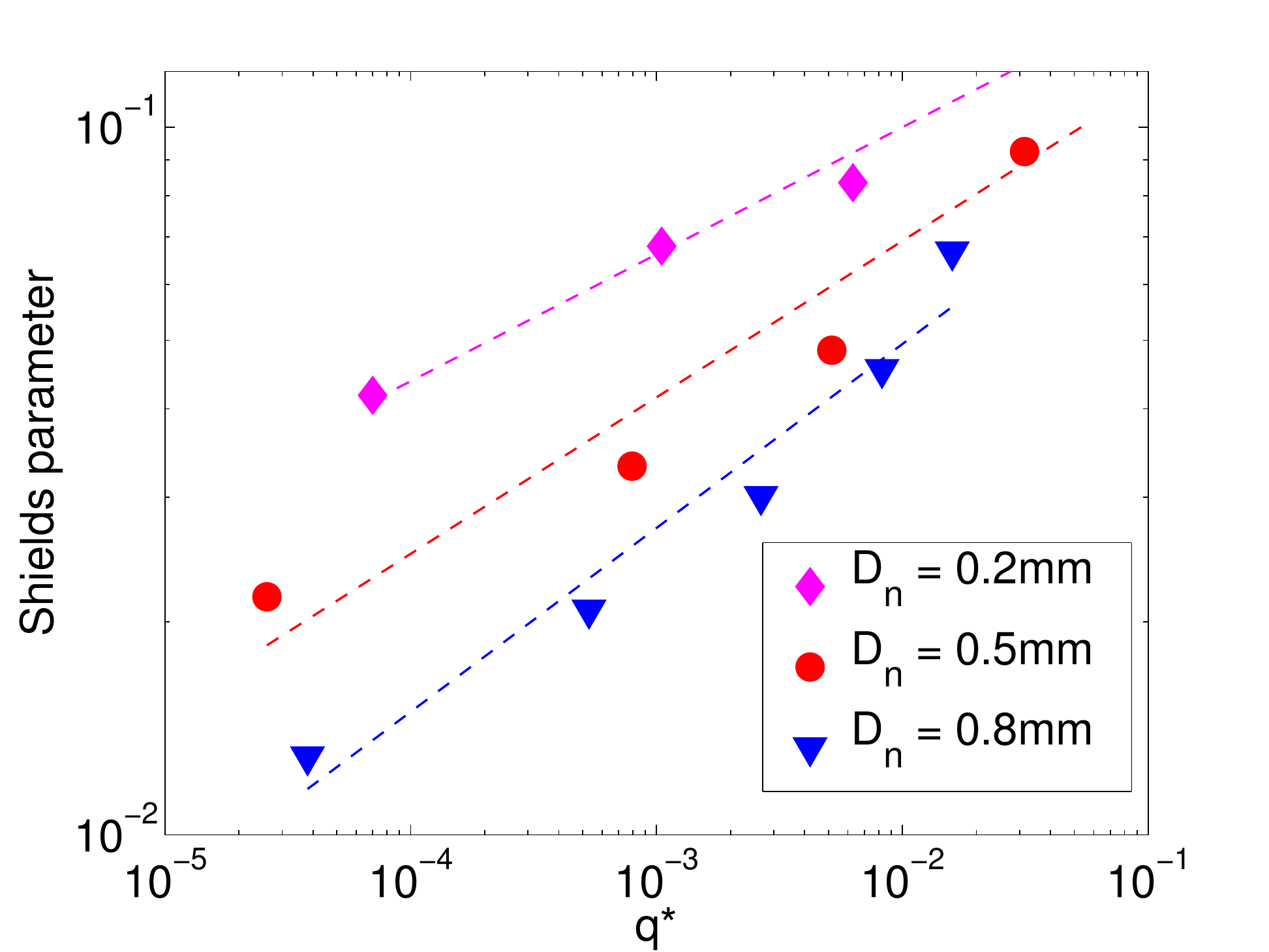}
  } 
  \vspace{0.05in}
  \subfloat[critical Shields parameter]{
  \includegraphics[width=0.495\textwidth]{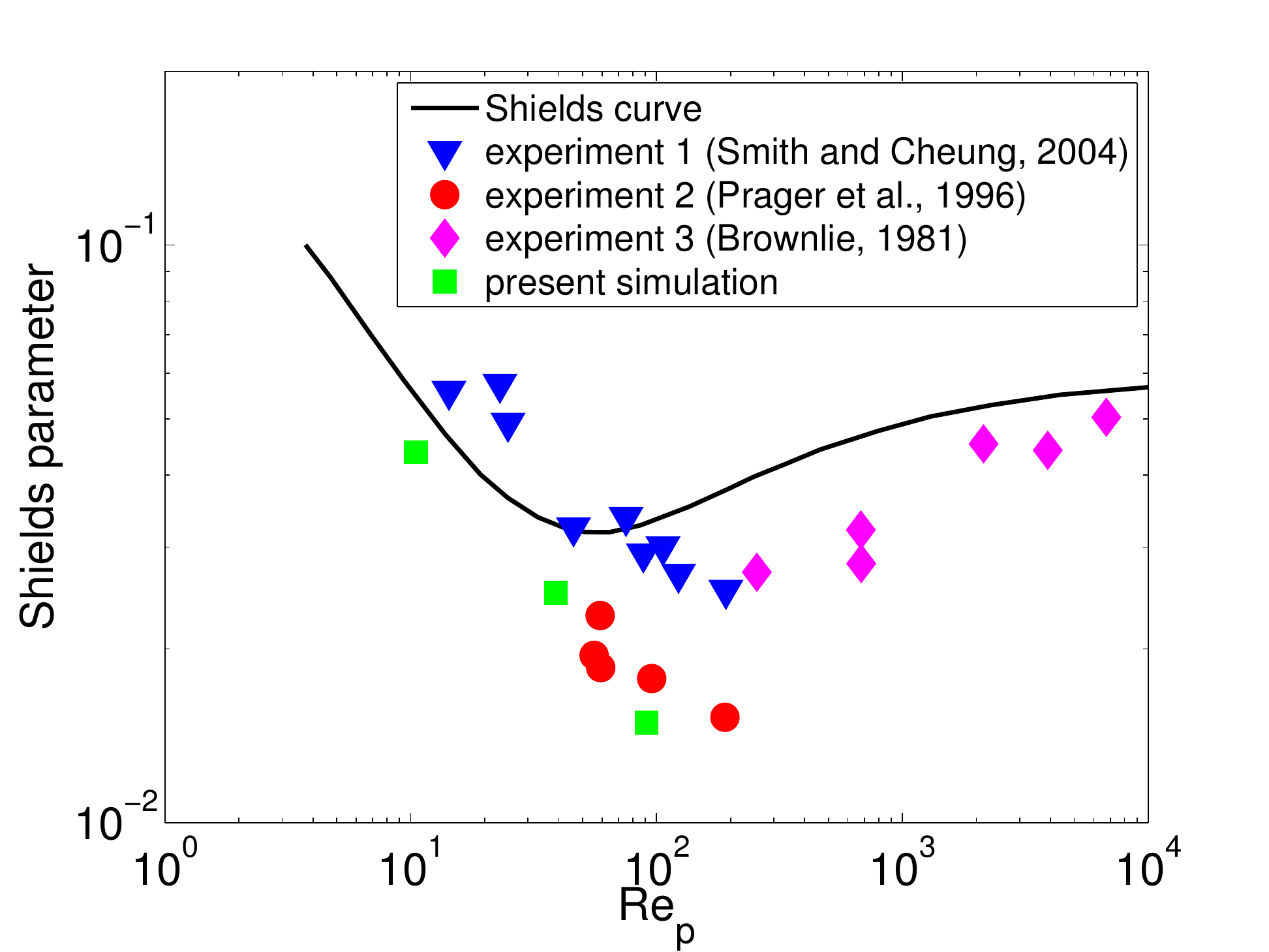}
  }
  \caption{Critical Shields parameters obtained in present simulations. The non-dimensional sediment
  transport rate is plotted as a function of Shields parameter in Panel (a). The critical Shields
  parameter plotted as a function of particle Reynolds number $Re_p$ in Panel (b).}
  \label{fig:criticalShields}
\end{figure}

The sediment transport rate plotted as a function of the Shields parameter is shown in
Fig.~\ref{fig:criticalShields}(a). It can be seen in the figure that transport rate increases with
the Shields parameter for irregular particles at different diameters. The sediment transport rates
of small particles are larger than those of larger particles, which is consistent with trend of the
experimental data obtained by~\cite{smith04im}. The critical Shields parameters obtained in the
simulations are shown in Fig.~\ref{fig:criticalShields}(b). From the comparison with the
experimental measurements~\citep{smith04im}, it can be seen that the proposed approach captures
decrease of the critical Shields stress due to the increase of the particle diameter. Moreover, the
results obtained in the CFD--DEM simulations are in the range of those obtained by experimental
measurements using carbonate sands~\citep{prager96exp}. The decrease of the critical Shields
parameter in the rough turbulent flow regime ($D_n$ = 0.8~mm, $Re_p$ = 80) reported
by~\cite{smith04im} is captured in the simulation. The present approach can capture the decrease of
critical Shields parameter because it can predict the increase of the particle drag and the rotation
due to particle irregularity, both of which contribute significantly to increase the chance of the
incipient motion.  On the other hand, the current simulations slightly under-predict the Shields
parameter in hydraulic smooth flow regime ($D_n$ = 0.2~mm, $Re_p$ = 10). From~\cite{smith04im}, the
Shields parameter of irregular particles in this flow regime should be slightly larger than that of
spherical particles, which is because the particle irregularity hinders the incipient motion. We
argue that the present approach only uses a few spherical particles to represent the irregular
particle in consideration of computational cost, and thus the geometric accuracy is not adequate to
capture the increase in critical Shields parameter.  However, since the under-prediction of the
proposed approach is very small, the overall agreement of the Shields parameters reported by the
present approach is satisfactory.

\subsection{Sediment Transport in a Periodic Channel}
\label{sec:periodic}

To test the capability of the proposed approach in the prediction of the sediment transport rate,
numerical simulations of sediment transport in a periodic channel are performed. Three
nominal diameters of the irregular particles used are 0.2~mm, 0.5~mm and 0.8~mm, which is consistent
with the incipient motion test in Section~\ref{sec:incipient}. The flow velocities
in the channel are higher than those in the incipient motion test at 0.4~m/s, 0.5~m/s, and 0.6~m/s.
This is because the flow velocities in the sediment transport experiments are higher than those used
in incipient motion experiments~\citep{smith04im,smith05tr}. The proportion of different types of
irregular particles is also according to the measurement in~\cite{smith03ef}. To demonstrate the
influence of the particle irregularity to sediment transport, simulations of spherical particle of
0.5~mm are also performed at the same velocities. 

\begin{figure}[htbp]
  \centering
  \includegraphics[width=0.75\textwidth]{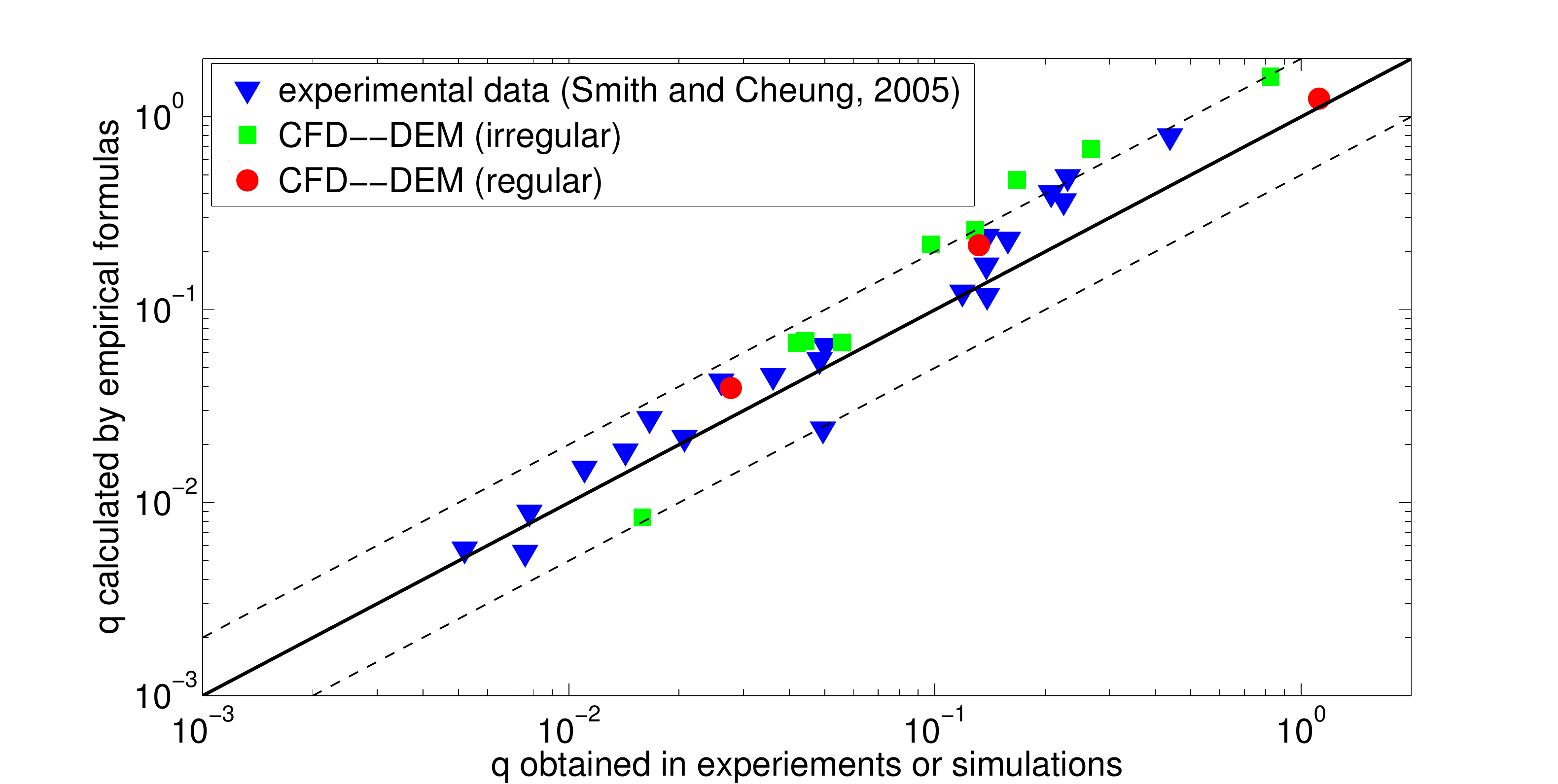}
  \caption{Comparison between the sediment transport rates obtained by using the regression formula
    in Eq.~(\ref{eq:gs-qt}) and those obtained in experimental measurements and numerical
    simulations.}
  \label{fig:transRates}
\end{figure}

Although the setup of the numerical simulations is according to the experimental measurements, the
predictions obtained by the numerical approach cannot compare directly to the experimental results.
This is because the detailed setup of the experiments in each run is unknown. Therefore, the results
from both numerical simulations and experimental measurements are plotted in comparison with the
empirical formula proposed by~\cite{ackers73st} for validation. The formula proposed
by~\cite{ackers73st} makes use of the grain-size parameter:
\begin{equation}
  D_* = D\left[ \frac{(s-1)g}{\nu^2} \right]^{1/3},
  \label{eq:gs-param}
\end{equation}
where $D = D_{n}$ is the characteristic grain-size parameter. The sediment transport rate $q$ is
defined as:
\begin{equation}
  q = B\left( \frac{F}{A} - 1 \right)^M D \bar{u}s\left( \frac{\bar{u}}{u_*} \right)^n,
  \label{eq:gs-qt}
\end{equation}
where $u_*$ is the friction velocity; $n$ is the transition parameter given by $n = 1 - 0.56
\log(D_*)$; $\bar{u}$ is depth-averaged flow velocity; $s = \rho_s/\rho_f$ is the sand specific
gravity. The coefficient $F$ denotes the sediment mobility given by:
\begin{equation}
  F = \frac{u_*^n\bar{u}^{1-n}}{\sqrt{gD(s-1)}[\sqrt{32}\log(10d/D)]^{1-n}}.
  \label{eq:gs-f}
\end{equation}
The initiation of motion is considered in coefficient $A$:
\begin{equation}
  A = \frac{0.23}{\sqrt{D_*}}+0.14.
  \label{eq:gs-a}
\end{equation}
The expressions for $B$ and $M$ are updated in~\cite{ackers93st}:
\begin{subequations}
 \label{eq:BM}
 \begin{align}
  \log(B) & = -3.46+2.79\log(D_*) - 0.98\left( \log(D_*) \right)^2, \\
  M & = \frac{6.83}{D_*}+1.67.
 \end{align}
\end{subequations}
The comparison between sediment transport rates obtained in the simulations and the predictions
from the empirical formula in Eq.~(\ref{eq:gs-qt}) is shown in Fig.~\ref{fig:transRates}. The solid
line plotted in the figure indicates perfect agreement; two dash lines of slope 0.5 and 2 indicate
under-prediction and over-prediction by a factor of two, respectively. It can be seen in the figure
that the results obtained by using spherical particles are consistent with the
predictions by the empirical formula. This shows the present CFD--DEM solver can capture the
averaged rate of spherical sediment transport in a periodic channel.  In addition, most results
obtained by using bonded particles are above the solid line, which is consistent with the
experimental measurements. This shows that the sediment transport rates of irregular sediment
particles are smaller than spherical particles.  Comparing the results obtained using both spherical
and irregular particles, the present modeling approach captures the influence of particle
irregularity which blocks the motion of the irregular particle. Therefore, the proposed algorithm to
model irregular sediment particles is capable of predicting bed load sediment transport of irregular
particles.

\section{Conclusion}
\label{sec:conclusion}

In this work we proposed a simple, efficient approach to represent common sediment grain shapes with
composite particles consisting of bonded spheres.  The fluid forces are computed and applied on each
sphere, which is in contrast to previous works that applied the forces to the center of mass of the
composite particle. The proposed approach better represents the fluid-exerted torque on the
composite particles, which are important for some forms of sediment transport mode (e.g., saltation)
in the bottom boundary layer. Moreover, this method is much simpler and computationally more
efficient than representing irregular shaped with non-spherical shapes such as cylinders,
ellipsoids, and polygons as pursed in previous works. The simplicity and efficiency are critical for
simulating large systems with many particles and offer the flexibility of modeling natural sediments
consisting of an arbitrary mixture of particle shapes.  Numerical simulations were performed to
demonstrate the merits and capability of the proposed method. Specifically, comparison of simulation
results with experiments showed that the current method is able to accurately predict the falling
characteristics, terminal velocity, threshold of incipient motion, and transport rate of natural
sediments. Therefore, the proposed method is a promising approach for faithful representation of
natural sediment, which leads to accurate simulations of their transport dynamics.

While this work has focused on non-cohesive sediments, the proposed method also opens the
possibility for first-principle-based simulation of the flocculation and sedimentation dynamics of
cohesive sediments. Elucidation of these physical mechanisms can provide much needed improvement on
the prediction capability and physical understanding of the muddy coasts evolution, which occupy a
large portion of the world's coastlines including tidal flats, slat marshes, lagoons, and
estuaries. The method will be applied to the settling and transport of cohesive sediments in future
work.

\section*{Acknowledgment}

The computational resources used for this project are provided by the Advanced Research Computing
(ARC) of Virginia Tech, which is gratefully acknowledged.  RS gratefully acknowledge partial funding
of graduate research assistantship from the Institute for Critical Technology and Applied Science
(ICTAS, grant number 175258) in this effort.

\section*{Reference}
\bibliographystyle{elsarticle-harv}
\bibliography{ACS-PRF,ACS-New,bonded-particles,extra_rui}

\begin{thebibliography}{49}
\expandafter\ifx\csname natexlab\endcsname\relax\def\natexlab#1{#1}\fi
\expandafter\ifx\csname url\endcsname\relax
  \def\url#1{\texttt{#1}}\fi
\expandafter\ifx\csname urlprefix\endcsname\relax\def\urlprefix{URL }\fi

\bibitem[{Ackers and White(1973)}]{ackers73st}
Ackers, P., White, W.~R., 1973. Sediment transport: New approach and analysis.
  Journal of the Hydraulics Division 99~(11), 2041--2060.

\bibitem[{Ackers and White(1993)}]{ackers93st}
Ackers, P., White, W.~R., 1993. Sediment transport in open channels: {Ackers
  and White} update. Proceedings of the Institution of Civil Engineers, Water,
  Maritime and Energy 101~(4), 247--249.

\bibitem[{Anderson and Jackson(1967)}]{anderson67}
Anderson, T., Jackson, R., 1967. A fluid mechanical description of fluidized
  beds: Equations of motion. Industrial and Chemistry Engineering Fundamentals
  6, 527--534.

\bibitem[{Ball and Melrose(1997)}]{ball97si}
Ball, R.~C., Melrose, J.~R., 1997. A simulation technique for many spheres in
  quasi-static motion under frame-invariant pair drag and {Brownian} forces.
  Physica A: Statistical Mechanics and its Applications 247~(1), 444--472.

\bibitem[{Calantoni et~al.(2004)Calantoni, Holland, and
  Drake}]{calantoni2004modelling}
Calantoni, J., Holland, K.~T., Drake, T.~G., 2004. Modelling sheet-flow
  sediment transport in wave-bottom boundary layers using discrete-element
  modelling. Philosophical Transactions-Royal Society Of London Series A:
  Mathematical Physical And Engineering Sciences 362, 1987--2002.

\bibitem[{Capecelatro and Desjardins(2013)}]{Capecelatro_13_AE}
Capecelatro, J., Desjardins, O., 2013. An {Euler--Lagrange} strategy for
  simulating particle-laden flows. Journal of Computational Physics 238, 1--31.

\bibitem[{Cundall and Strack(1979)}]{cundall79}
Cundall, P., Strack, D., 1979. A discrete numerical model for granular
  assemblies. G\'{e}otechnique 29, 47--65.

\bibitem[{Demir(2000)}]{demir00if}
Demir, T., 2000. The influence of particle shape on bedload transport in
  coarse-bed river channels. Ph.D. thesis, Durham University.

\bibitem[{Di~Felice(1994)}]{di1994voidage}
Di~Felice, R., 1994. The voidage function for fluid-particle interaction
  systems. International Journal of Multiphase Flow 20~(1), 153--159.

\bibitem[{Favier et~al.(1999)Favier, Abbaspour-Fard, Kremmer, and
  Raji}]{favier1999shape}
Favier, J.~F., Abbaspour-Fard, M.~H., Kremmer, M., Raji, A.~O., 1999. Shape
  representation of axi-symmetrical, non-spherical particles in discrete
  element simulation using multi-element model particles. Engineering
  Computations 16~(4), 467--480.

\bibitem[{Guo et~al.(2013)Guo, Wassgren, Hancock, Ketterhagen, and
  Curtis}]{guo2013granular}
Guo, Y., Wassgren, C., Hancock, B., Ketterhagen, W., Curtis, J., 2013. Granular
  shear flows of flat disks and elongated rods without and with friction.
  Physics of Fluids (1994-present) 25~(6), 063304.

\bibitem[{Guo et~al.(2012)Guo, Wassgren, Ketterhagen, Hancock, James, and
  Curtis}]{guo2012numerical}
Guo, Y., Wassgren, C., Ketterhagen, W., Hancock, B., James, B., Curtis, J.,
  2012. A numerical study of granular shear flows of rod-like particles using
  the discrete element method. Journal of Fluid Mechanics 713, 1--26.

\bibitem[{Haider and Levenspiel(1989)}]{haider89dc}
Haider, A., Levenspiel, O., 1989. Drag coefficient and terminal velocity of
  spherical and nonspherical particles. Powder technology 58~(1), 63--70.

\bibitem[{Ikeguchi(2004)}]{ikeguchi04prb}
Ikeguchi, M., 2004. Partial rigid-body dynamics in {NPT}, {NPAT} and
  {NP$\gamma$T} ensembles for proteins and membranes. Journal of computational
  chemistry 25~(4), 529--541.

\bibitem[{Issa(1986)}]{issa86so}
Issa, R.~I., 1986. Solution of the implicitly discretised fluid flow equations
  by operator-splitting. Journal of Computational Physics 62~(1), 40--65.

\bibitem[{Kafui et~al.(2002)Kafui, Thornton, and Adams}]{kafui02}
Kafui, K., Thornton, C., Adams, M., 2002. Discrete particle--continuum fluid
  modelling of gas--solid fluidised beds. Chemical Engineering Science 57~(13).

\bibitem[{Kempe et~al.(2014)Kempe, Vowinckel, and Fr\"{o}hlich}]{kempe14ot}
Kempe, T., Vowinckel, B., Fr\"{o}hlich, J., 2014. On the relevance of collision
  modeling for interface-resolving simulations of sediment transport in open
  channel flow. International Journal of Multiphase Flow 58, 214--235.

\bibitem[{Kodam et~al.(2009)Kodam, Bharadwaj, Curtis, Hancock, and
  Wassgren}]{kodam2009force}
Kodam, M., Bharadwaj, R., Curtis, J., Hancock, B., Wassgren, C., 2009. Force
  model considerations for glued-sphere discrete element method simulations.
  Chemical Engineering Science 64~(15), 3466--3475.

\bibitem[{Kodam et~al.(2010{\natexlab{a}})Kodam, Bharadwaj, Curtis, Hancock,
  and Wassgren}]{kodam2010cylindrical-a}
Kodam, M., Bharadwaj, R., Curtis, J., Hancock, B., Wassgren, C.,
  2010{\natexlab{a}}. Cylindrical object contact detection for use in discrete
  element method simulations. {Part I}--contact detection algorithms. Chemical
  Engineering Science 65~(22), 5852--5862.

\bibitem[{Kodam et~al.(2010{\natexlab{b}})Kodam, Bharadwaj, Curtis, Hancock,
  and Wassgren}]{kodam2010cylindrical-b}
Kodam, M., Bharadwaj, R., Curtis, J., Hancock, B., Wassgren, C.,
  2010{\natexlab{b}}. Cylindrical object contact detection for use in discrete
  element method simulations, {Part II}--experimental validation. Chemical
  Engineering Science 65~(22), 5863--5871.

\bibitem[{Kodam et~al.(2012)Kodam, Curtis, Hancock, and
  Wassgren}]{kodam2012discrete}
Kodam, M., Curtis, J., Hancock, B., Wassgren, C., 2012. Discrete element method
  modeling of bi-convex pharmaceutical tablets: Contact detection algorithms
  and validation. Chemical Engineering Science 69~(1), 587--601.

\bibitem[{Kruggel-Emden et~al.(2008)Kruggel-Emden, Rickelt, Wirtz, and
  Scherer}]{kruggel2008study}
Kruggel-Emden, H., Rickelt, S., Wirtz, S., Scherer, V., 2008. A study on the
  validity of the multi-sphere discrete element method. Powder Technology
  188~(2), 153--165.

\bibitem[{Krumbein(1941)}]{krumbein1941measurement}
Krumbein, W.~C., 1941. Measurement and geological significance of shape and
  roundness of sedimentary particles. Journal of Sedimentary Research 11~(2).

\bibitem[{Miller~III et~al.(2002)Miller~III, Eleftheriou, Pattnaik, Ndirango,
  Newns, and Martyna}]{miller02sqs}
Miller~III, T., Eleftheriou, M., Pattnaik, P., Ndirango, A., Newns, D.,
  Martyna, G., 2002. Symplectic quaternion scheme for biophysical molecular
  dynamics. The Journal of chemical physics 116~(20), 8649--8659.

\bibitem[{O'Connor et~al.(1997)O'Connor, Torczynski, Preece, Klosek, and
  Williams}]{connor97}
O'Connor, R., Torczynski, J., Preece, D., Klosek, J., Williams, J., 1997.
  Discrete element modeling of sand production. International Journal of Rock
  Mechanics and Mining Sciences 34~(3-4), 1--15.

\bibitem[{{OpenCFD}(2016)}]{openfoam}
{OpenCFD}, 2016. OpenFOAM User Guide. See also
  \url{http://www.opencfd.co.uk/openfoam}.

\bibitem[{Plimpton(1995)}]{lammps}
Plimpton, J., 1995. Fast parallel algorithms for short-range molecular
  dynamics. Journal of Computational Physics 117, 1--19, see also
  \url{http://lammps.sandia.gov/index.html}.

\bibitem[{Prager et~al.(1996)Prager, Southard, and
  Vivoni-Gallart}]{prager96exp}
Prager, E.~J., Southard, J.~B., Vivoni-Gallart, E.~R., 1996. Experiments on the
  entrainment threshold of well-sorted and poorly sorted carbonate sands.
  Sedimentology 43~(1), 33--40.

\bibitem[{Price et~al.(2007)Price, Murariu, and Morrison}]{price2007sphere}
Price, M., Murariu, V., Morrison, G., 2007. Sphere clump generation and
  trajectory comparison for real particles. In: Proceedings of Fifth
  International Conference on Discrete Element Methods. August 27--29,
  Brisbane, Australia.

\bibitem[{Rusche(2003)}]{rusche03co}
Rusche, H., 2003. Computational fluid dynamics of dispersed two-phase flows at
  high phase fractions. Ph.D. thesis, Imperial College London (University of
  London).

\bibitem[{Saffman(1965)}]{saffman65th}
Saffman, P., 1965. The lift on a small sphere in a slow shear flow. Journal of
  fluid mechanics 22~(02), 385--400.

\bibitem[{Schmeeckle(2014)}]{schmeeckle14ns}
Schmeeckle, M.~W., 2014. Numerical simulation of turbulence and sediment
  transport of medium sand. Journal of Geophysical Research: Earth Surface 119,
  1240--1262.

\bibitem[{Smith(2003)}]{smith03ef}
Smith, D.~A., 2003. Effect of particle shape on grain size, hydraulic, and
  transport characteristics of calcareous sand. Ph.D. thesis, University of
  Hawaii at Manoa.

\bibitem[{Smith and Cheung(2003)}]{smith03sc}
Smith, D.~A., Cheung, K.~F., 2003. Settling characteristics of calcareous sand.
  Journal of Hydraulic Engineering 129~(6), 479--483.

\bibitem[{Smith and Cheung(2004)}]{smith04im}
Smith, D.~A., Cheung, K.~F., 2004. Initiation of motion of calcareous sand.
  Journal of Hydraulic Engineering 130~(5), 467--472.

\bibitem[{Smith and Cheung(2005)}]{smith05tr}
Smith, D.~A., Cheung, K.~F., 2005. Transport rate of calcareous sand in
  unidirectional flow. Sedimentology 52~(5), 1009--1020.

\bibitem[{Song et~al.(2006)Song, Turton, and Kayihan}]{song2006contact}
Song, Y., Turton, R., Kayihan, F., 2006. Contact detection algorithms for {DEM}
  simulations of tablet-shaped particles. Powder Technology 161~(1), 32--40.

\bibitem[{Sun et~al.(2007)Sun, Battaglia, and Subramaniam}]{sun07ht}
Sun, J., Battaglia, F., Subramaniam, S., 2007. Hybrid two-fluid {DEM}
  simulation of gas-solid fluidized beds. Journal of Fluids Engineering
  129~(11), 1394--1403.

\bibitem[{Sun and Xiao(2015{\natexlab{a}})}]{sun14db2}
Sun, R., Xiao, H., 2015{\natexlab{a}}. Diffusion-based coarse graining in
  hybrid continuum--discrete solvers: {Applications} in {CFD--DEM}.
  International Journal of Multiphase Flow 72, 233--247.

\bibitem[{Sun and Xiao(2015{\natexlab{b}})}]{sun14db1}
Sun, R., Xiao, H., 2015{\natexlab{b}}. Diffusion-based coarse graining in
  hybrid continuum--discrete solvers: Theoretical formulation and a priori
  tests. International Journal of Multiphase Flow 77, 142 -- 157.

\bibitem[{Sun and Xiao(2016{\natexlab{a}})}]{sun16awr}
Sun, R., Xiao, H., 2016{\natexlab{a}}. {CFD--DEM} simulations of
  current-induced dune formation and morphological evolution. Advances in Water
  Resources 92, 228 -- 239.

\bibitem[{Sun and Xiao(2016{\natexlab{b}})}]{sun2016sedi}
Sun, R., Xiao, H., 2016{\natexlab{b}}. {SediFoam}: A general-purpose,
  open-source {CFD-DEM} solver for particle-laden flow with emphasis on
  sediment transport. Computers and Geosciences 89, 207--b219.

\bibitem[{Syamlal et~al.(1993)Syamlal, Rogers, and O'Brien}]{mfix93}
Syamlal, M., Rogers, W., O'Brien, T., 1993. {MFIX} documentation: Theory guide.
  Tech. rep., National Energy Technology Laboratory, Department of Energy, see
  also URL \url{http://www.mfix.org}.

\bibitem[{Tran-Cong et~al.(2004)Tran-Cong, Gay, and Michaelides}]{tran2004drag}
Tran-Cong, S., Gay, M., Michaelides, E.~E., 2004. Drag coefficients of
  irregularly shaped particles. Powder Technology 139~(1), 21--32.

\bibitem[{Tsuji et~al.(1993)Tsuji, Kawaguchi, and Tanaka}]{tsuji93}
Tsuji, Y., Kawaguchi, T., Tanaka, T., 1993. Discrete particle simulation of
  two-dimensional fluidized bed. Powder Technolgy 77~(79-87).

\bibitem[{van Rijn(1984)}]{rijn84se1}
van Rijn, L., 1984. Sediment transport, part {I}: Bed load transport. Journal
  of hydraulic engineering 110~(10), 1431--1456.

\bibitem[{Wen and Yu(1966)}]{wen1966mechanics}
Wen, C., Yu, Y., 1966. Mechanics of fluidization. The Chemical Engineering
  Progress Symposium Series 162, 100--111.

\bibitem[{Yoshizawa and Horiuti(1985)}]{yoshizawa85sd}
Yoshizawa, A., Horiuti, K., 1985. A statistically-derived subgrid-scale kinetic
  energy model for the large-eddy simulation of turbulent flows. Journal of the
  Physical Society of Japan 54~(8), 2834--2839.

\bibitem[{Zingg(1935)}]{zingg1935beitrag}
Zingg, T., 1935. Beitrag zur schotteranalyse. Ph.D. thesis, Diss. Naturwiss.
  ETH Z{\"u}rich, Nr. 849.

\end{thebibliography}

\appendix
\setcounter{secnumdepth}{0}
\section{Appendix}

\renewcommand\thefigure{A.\arabic{figure}}
\renewcommand\theequation{A.\arabic{equation}}
\setcounter{figure}{0}

\section{Representation of Equant Particles}

  The long, intermediate, and short axes of equant particles have similar length ($\frac{2}{3}D_l <
  D_i < D_l$ and $\frac{2}{3}D_i < D_s < D_i$). In the proposed method, we attempt to use the
  smallest number of component spheres to represent each particle and maximize the wetted surface
  for each component sphere. To this end, the equant particles are further classified to four
  subtypes, each represented with one main sphere and zero to three auxiliary spheres depending on
  the particle geometry:
  \begin{enumerate}
  \item Loosely written as $D_s \approx D_i \approx D_l$, the three axes have almost equal lengths
    as defined by $(1-\varepsilon) D_l < D_i < D_l$ and $(1-\varepsilon)D_i < D_s < D_i$ with
    $\varepsilon = 0.1$, and thus the particle can be considered close to spherical.  A single
    component sphere with diameter $d_p = (D_l \, D_i \, D_s)^{1/3}$ is used to represent the
    particle. See Fig.~\ref{fig:layout-table}(d.1).

  \item Loosely written as $D_s \approx D_i < D_l$, the short and intermediate axes have almost
    equal lengths ($(1-\varepsilon) D_i < D_s < D_i$) but the longer axis is much longer than the
    intermediate axis ($\frac{2}{3}D_l < D_i < (1-\varepsilon)D_l$).  A larger main component sphere
    of diameter $d_{p, main} = \sqrt{D_s D_i}$ and a smaller auxiliary component sphere of diameter
    $d_{p, auxi} = D_l - d_{p,main}$ are used to represent the particle. See
    Fig.~\ref{fig:layout-table}(d.2).

  \item Loosely written as $D_s < D_i \approx D_l$, the long and intermediate axes have almost equal
    lengths ($(1-\varepsilon) D_l < D_i < D_l$) but the short axis is shorter than the intermediate
    axis ($\frac{2}{3}D_i < D_s < (1-\varepsilon)D_i$).  A larger main component sphere with $d_{p,
    main} = D_s$ and two small auxiliary component spheres with $d_{p, auxi} = D_l -
    D_s$ are used to represent the particle. The angle between the auxiliary particles is $\theta_e =
    2\sin^{-1}((D_i+D_s-D_l)/D_l)$. See Fig.~\ref{fig:layout-table}(d.3).

  \item Loosely written as $D_s < D_i < D_l$, neither long and intermediate axes nor intermediate
    and short axes have almost equal lengths, i.e., ($\frac{2}{3}D_l < D_i < (1-\varepsilon) D_l $)
    and axis ($\frac{2}{3}D_i < D_s < (1-\varepsilon)D_i$). A larger main component sphere with
    $d_{p, main} = D_s$ and three small auxiliary component spheres with $d_{p, auxi} = 0.5(D_l
    - D_s)$ are used to represent the particle. The arrangement of the sediment particles is shown
    in Fig.~\ref{fig:layout-table}(d.4). The main component is aligned with two auxiliary components
    on both sizes, while angle between the third auxiliary components and the first one is $\theta_e
    = 2\sin^{-1}((2D_i+D_s-D_l)/(D_l+D_s))$.
  \end{enumerate}
\end{document}